\documentclass[12pt]{article}

\usepackage{amsmath}
\usepackage{graphicx}
\usepackage{enumerate}
\usepackage{url} % not crucial - just used below for the URL
\usepackage{amsfonts}%
\usepackage{amssymb}
\usepackage{color}
\usepackage{float,subfigure}
\usepackage[round]{natbib}
\usepackage{setspace}
\usepackage{comment}
\usepackage{verbatim}
\usepackage{rotating}
\usepackage{microtype}
\usepackage[normalem]{ulem}
\usepackage{url}
\usepackage[hidelinks]{hyperref} %,bookmarks=FALSE
\usepackage{cancel}
\usepackage{animate}
\usepackage{mathtools}
\usepackage{epsfig}
\usepackage{epstopdf}
\usepackage{amsthm}
\usepackage{xcolor} %gives more colors
\usepackage{chngcntr}
\usepackage{xr-hyper} % cross reference between the main text and appendix

\newcommand{\blind}{1}

\newcommand{\blam}{ \mbox{\boldmath $ \lambda $} }

\newcommand{\bbeta}{ \ensuremath{\boldsymbol{\beta}}}

\newcommand{\balpha}{ \mbox{\boldmath $ \alpha $} }

\newcommand{\btheta}{ \mbox{\boldmath $ \theta $} }

\newcommand{\bmu}{ \mbox{\boldmath $\mu$} }

\newcommand{\bSig}{ \mbox{\boldmath $\Sigma$} }

\newcommand{\sig}{ \ensuremath{\sigma}}

\newcommand{\lam}{ \ensuremath{\lambda}}

\newcommand{\bPhi}{ \mbox{\boldmath $\Phi$} }

\newcommand{\bPsi}{ \mbox{\boldmath $\Psi$} }

\newcommand{\bgam}{ \mbox{\boldmath $\gamma$} }
\newcommand{\bgamma}{ \mbox{\boldmath $\gamma$} }

\newcommand{\bzero}{\textbf{0}}

\newcommand{\bD}{ {\bf D} }

\newcommand{\bG}{ {\bf G} }

\newcommand{\bW}{ {\bf W} }
\newcommand{\bx}{ {\bf x} }

\newcommand{\bz}{ {\bf z} }

\newcommand{\given}{\,\vert\,}

\newcommand{\CAR}{\mbox{$\text{CAR}$}}
\newcommand{\MCAR}{\mbox{$\text{MCAR}$}}

\newcommand{\IW}{\mbox{$\text{InvWish}$}}
\newcommand{\IG}{\mbox{$\text{IG}$}}

\newcommand{\Pois}{\mbox{$\text{Pois}$}}

\newcommand{\Gam}{\mbox{$\text{Gamma}$}}
\newcommand{\N}{\mbox{Norm}}

\usepackage[dvipsnames]{xcolor}
\begin{document}

\def\spacingset#1{\renewcommand{\baselinestretch}%
{#1}\small\normalsize} \spacingset{1}

%%%%%%%%%%%%%%%%%%%%%%%%%%%%%%%%%%%%%%%%%%%%%%%%%%%%%%%%%%%%%%%%%%%%%%%%%%%%%%

\if1\blind
{
  \title{\bf Restricted Multivariate Spatial Modeling}
  \author{Jihyeon Kwon\\
    Department of Epidemiology and Biostatistics, Drexel University\\
    and \\
    Harrison Quick\thanks{
    quic0038@umn.edu}\hspace{.2cm}\\
    Division of Biostatistics and Health Data Science, University of Minnesota}
  \maketitle
  \thispagestyle{empty}    % suppress number on title page
  \setcounter{page}{0}     % reset counter so next page becomes 1
} \fi

\if0\blind
{
  \bigskip
  \bigskip
  \bigskip
  \begin{center}
    {\LARGE\bf Restricted Multivariate Spatial Modeling}
\end{center}
  \medskip
} \fi

\bigskip
\begin{abstract}
When modeling health events in small areas, the conditional autoregressive (CAR) framework of Besag, York, and Molli\'{e} (BYM) is widely used. For multiple outcomes, the multivariate CAR (MCAR) extension accommodates dependence among diseases that share risk factors, in addition to spatial dependence, and can also jointly model demographic subgroups for a single disease, allowing information to be borrowed across related populations. However, recent studies have shown that the BYM CAR model can be overly informative, leading to excessively precise estimates. While the MCAR model is expected to be more informative due to additional information shared across subgroups, its level of informativeness has not been previously quantified. We propose a framework to measure MCAR model informativeness as an extension of prior work and introduce a method to control it, ensuring the model contributes comparably to each subgroup. We achieve this through a reparameterization of the MCAR model within a computationally efficient framework. We demonstrate how the MCAR model compares with the BYM CAR model in terms of informativeness and oversmoothing and highlight the advantages of the restricted MCAR model using county-level heart disease death data stratified by race and sex.
\end{abstract}

\noindent%
{\it Keywords:}  Spatial Model, Oversmoothing, Disease Mapping, Multivariate Model, Informativeness
% 3 to 6 keywords, that do not appear in the title
\vfill

\newpage
\spacingset{1.9} % DON'T change the spacing!

%%%%%%%%%%%%%%%%%%%%%%%%%%%%%%%%%%%%%%%%%%%%%%%%%%%%%%%%%%%%%%%%%%%%%%%%%%%%%%%%

\section{Introduction}\label{sec:intro}

Disease mapping can help epidemiologists and public health researchers better visualize---and understand---the geographical distribution of diseases. For instance, early work in the field of spatial statistics by \citet{clayton:kaldor} used random effects with a conditional autoregressive (CAR) model \citep{besag74} to make inference on county-level rates of lip cancer in Scotland by borrowing strength across neighboring counties.
%developed models with spatial random effects based on adjacency structures that
%{\color{orange}while accounting for spatial correlations, which helps to address the issue of unstable crude rates}.
Shortly thereafter,
%Following the seminal work of \citet{besag74}, the conditional autoregressive (CAR) model framework of
Besag, York, and Molli\'{e} (BYM) published their seminal work \citep{bym} which proposed a fully Bayesian extension of the CAR model of \citet{besag74} that included both spatial and non-spatial random effects to explore geographic trends in cancer in France and the North of England.  While alternative approaches for modeling health outcome data based on the BYM CAR model framework have been proposed in recent years \citep[e.g.,][]{leroux:car,riebler:intuitive2016,datta:dagar}, the BYM CAR model remains ubiquitous in the fields of disease mapping and spatial epidemiology.

%Since the seminal work by \cite{besag74} and \cite{bym}, the Besag-York-Molli\'{e} (BYM) conditional autoregressive (CAR) model has been {\color{orange}foundational} in disease mapping literature. {\color{orange}Its core contribution lies in providing a principled structure that separates} spatially structured and unstructured heterogeneity, improving inference for regions with small populations and enabling more reliable quantification of geographic distribution in disease risk.

Implicit in the design of these various disease mapping strategies is the desire to leverage dependencies in the data to improve estimates of rates for spatial regions with small counts.  
Moving beyond mere \emph{spatial} dependencies, early approaches utilized \emph{univariate} spatial models to analyze inherently \emph{multivariate} data \citep[e.g.,][]{waller:Hierarchical1997,knorr-held:besag,knorr-held:2000,knorr-held:best} or focused on simple bivariate structures \citep[e.g.,][]{kim:Bivariate2001} before \citet{gelfand:Proper2003} and \citet{carlin:banerjee:2003} extended and expanded upon the theoretical work of \citet{mardia:Multidimensional1988} to develop fully Bayesian multivariate CAR (MCAR) models to jointly model two or more spatially referenced outcomes. While further enhancements in the area of multivariate spatial modeling have been developed in the years that followed \citep[e.g.,][]{jin:Generalized2005,JBC,m-b:2013,botella-rocamora:unifying2015}, in this paper we will focus our attention on an improper version of the MCAR model of \citet{gelfand:Proper2003}. %jin:Generalized2005, , \citet{martinez:beneito}, and other

While the pursuit of more precise estimates is valid, the notion that estimates can be \emph{too} precise has been a topic of recent discussion.  For instance, %early work by
%While spatial models are valuable for modeling and/or accounting for spatial autocorrelation, recent studies have drawn attention to potential drawbacks.
%\citet{bernardinelli} examined hyperprior sensitivity, and
\citet{duncan:Spatial2017} compared the effect that different spatial weight matrices had on how spatially smooth the resulting rate estimates were, whereas \citet{duncan:Comparing2020} developed a ``goodness-of-smoothing'' measure to assess under- and oversmoothing, noting that smoothness and precision tend to be interconnected.
%{\color{red}explored spatial smoothing}, noting that over- and under-smoothing in spatial models remain understudied concerns.
\citet{bym:info} took this idea one step further by quantifying the weight given to the model relative to the weight given to the data in the Bayesian framework---referred to as the model's \emph{informativeness}---and demonstrated via an analysis of Poisson-distributed %county-level heart disease death counts
death data that the BYM CAR model often produced estimates which put an inordinate amount of weight on the model's spatial structure; moreover, they provided guidance for how to \emph{restrict} the model's informativeness.  Follow-up work by \citet{song:bin} and \citet{quick:reliable} extended this framework to the analysis of binomially distributed data and introduced criteria for restricting the model's informativeness based on thresholds for declaring estimates of rates as ``reliable'', respectively.
% than the actual observed data; \citet{bym:info} expressed this ``weight'' as the \emph{informativeness} of the model.  Additional work by the same authors \citep{song:bin} then extended this framework to the analysis of binomially distributed data.
% 's contribution to the posterior, relative to the amount of information from observed data, is often much larger than expected, causing posterior distributions for region with small population to resemble the overall average, leading to oversmoothing.
%Moreover, based on the work of \citet{leroux:car}, \citet{datta:dagar}, and \citet{riebler:intuitive2016}---all of which proposed alternative frameworks for modeling spatial data that yielded estimates comparable to those obtained from the CAR model---it stands to reason that the conclusions of \citet{bym:info} are not \emph{unique} to the CAR model. % Prior studies have proposed methods to measure the informativeness of spatial models for Poisson data \citep{quick_evaluating_2021} and Binomial data \citep{song_restricted_2024}.

Building off of this existing work, the motivation of this paper is to investigate how the drawbacks identified in univariate spatial models are exacerbated in multivariate spatial settings, where more sources of information are used. % for borrowing.
\begin{comment}
For instance, consider a bivariate normal distribution where,
\begin{align*}
  \btheta =
  \begin{pmatrix}
  \theta_1 \\
  \theta_2
  \end{pmatrix}
  \sim
  \N \!\left(
  \bmu,
  \bSig
  \right),
  \qquad
  \bmu =
  \begin{pmatrix}
  \mu_1 \\
  \mu_2
  \end{pmatrix},
  \qquad
  \bSig =
  \begin{pmatrix}
  \sig_1^2 & \rho\,\sig_1\sig_2 \\
  \rho\,\sig_1\sig_2 & \sig_2^2
  \end{pmatrix}.
\end{align*}
Here, the conditional variance $V \left[\theta_2 \given \theta_1 \right] = \sig^2_2 \left(1 - \rho^2 \right)$ is strictly smaller than the marginal variance $V \left[ \theta_2 \right] = \sig^2_2$ if $\rho \ne 0$. This means incorporating correlated data increases the precision. However, when applied as a prior distribution, this increased precision may also aggravate the model informativeness, introducing a stronger influence from the prior.
\end{comment}
Specifically,
%In the MCAR models, although incorporating spatial correlations among different outcomes has shown clear benefits, the complex structure can obscure the level of informativeness contributed by the model. Thus,
our primary goals are to (a) quantify the informativeness of MCAR models and (b) propose a strategy for deliberately and efficiently restricting their informativeness.
In doing so, we aim to both raise awareness of oversmoothing in spatial and multivariate spatial data analyses and facilitate a more disciplined approach for researchers to analyze their data.
%allowing us to leverage multivariate dependencies effectively while mitigating the risks of excessive informativeness. Ultimately, we aim to develop a restricted MCAR model particularly suited for Poisson-distributed data.
To that end, this article will proceed as follows.
%This article is structured as follows.
Section~\ref{sec:methods} discusses our framework for assessing model informativeness in the MCAR model, and Section~\ref{sec:res-mcar} presents an alternative parameterization that facilitates efficient restrictions on the model's informativeness. %Section~\ref{sec:simul} demonstrates the quantification of informativeness through simulation studies.
In Section~\ref{sec:analysis}, we apply these methods to analyze county-level ischemic heart disease mortality rates by race and sex. Finally, we conclude with a discussion of the findings and future considerations in Section~\ref{sec:conclusion} .

\section{Multivariate Model Informativeness} \label{sec:methods}

We let $y_{ik}$ denote the number of events in region $i$ for group $k$ within a population of size $n_{ik}$ and assume $y_{ik} \sim \text{Pois} \left( n_{ik} \lambda_{ik} \right)$, where $\lambda_{ik}$ denotes the underlying event rate, for $i = 1, \ldots, I$ and $k = 1, \ldots, K$, following the recommendation of \citet{brillinger}. To address the challenge of quantifying model informativeness in this context, we begin by reviewing the relevant work of \citet{bym:info} in Section~\ref{sec:car review}, which serves as the basis for our approach. Building on these insights, we then introduce our framework for quantifying the informativeness of the MCAR model in Section~\ref{sec:mcar info}.

% \section{\color{red} Evaluation of the Multivariate Models' Informativeness}
% {\color{red}[I'd include a short ``$y_{ik}$ denotes... where $n_{ik}$... and $\lambda_{ik}$...'' paragraph here.  That will allow you to introduce the notation, and then you can have a ``\emph{Before we discuss our framework for quantifying and restricting the informativeness of the MCAR in Section~X.X, we will briefly summarize the work of [bym:info] in Section~X.X to establish a foundation for our work} sentence to set things up.  You can/should still have the ``This paper is organized as follows'' paragraph in the Intro, but \emph{this} paragraph will allow \emph{that} paragraph to be shorter.]}
% %\section{Informativeness of the Bayesian Spatial Models}

\subsection{Review of Model Informativeness}\label{sec:car review}

% {\color{lightgray}Recent work by \citet{quick_evaluating_2021} and \citet{song_restricted_2024} has highlighted the informativeness of the BYM CAR model in univariate settings.} For example, suppose $y_i \sim \Pois(n_i \lambda_i)$ \citep{brillinger_biometrics_1986}, where $n_i$ is the population size of region $i$ and $\lambda_i$ is the mortality rate per unit population.

To illustrate our approach for quantifying model informativeness, we temporarily assume $K = 1$, and for clarity, we start with a simple scenario before addressing more complex cases; i.e., we assume $\lam_{i} \sim \Gam(a_{i}, b_{i})$, yielding a posterior distribution of the form
\begin{align}
    \lam_{i} \given y_{i}, a_{i}, b_{i} \sim \Gam(a_{i} + y_{i}, b_{i} + n_{i}).
    \label{eq:lam_post_gamma}
\end{align}
Here, %{\color{blue}since $y_i$ and $n_i$ represent the number of events and the population size contributed by the data}, 
$a_{i}$ and $b_{i}$ can be interpreted as the number of events and the population size contributed by the prior, respectively.
%Similarly, $a_{i} + y_{i}$ and $b_{i} + n_{i}$ can be interpreted as the total number of events and population size contained in the posterior.
Moreover, since the second parameter of a gamma distribution is merely a scaling factor---e.g., if $\lam \sim \Gam \left(a, b\right)$ then $b \lam \sim \Gam \left(a, 1\right)$---this allows us to interpret $a_{i} + y_{i}$ as the total amount of information in the posterior distribution, where $y_{i}$ represents the amount of information contributed by the data and $a_{i}$ corresponds to the amount of information contributed by the prior.  %{\color{blue}As will be discussed in Section~\ref{sec:car_info}, \citet{bym:info} extended this framework to lognormal prior specifications and drew parallels to}  the concept of ``effective sample size'' in the Bayesian clinical trials literature (e.g., \citet{morita:2008}).

While assuming $\lam_{i} \sim \Gam \left(a_{i}, b_{i} \right)$ is convenient for demonstrating the separate contributions of the data and the prior, in practice, the disease mapping literature commonly uses a lognormal specification to model rate parameters, assuming models of the form
\begin{align}
  \theta_{i} = \log \lam_{i} \sim \N \left( \mu_{i}, \tau^2_{i} \right).
  \label{eq:ln}
\end{align}
Since the lognormal prior in \eqref{eq:ln} is not conjugate, \citet{bym:info} suggested approximating its informativeness by equating the mean and variance of the gamma and lognormal priors. This yields $\tau^2_{i} = \log \left( 1 / a_{i} + 1 \right)$ and $\mu_{i} = \log \left( a_{i} / b_{i} \right) - \tau^2_{i} / 2$, providing an approximation for the informativeness of the lognormal model as follows:
\begin{align}
  \widehat{a}_{i} = \frac{1}{\exp \tau^2_{i} - 1}. \label{eq:info_lognorm}
\end{align}

\citet{bym:info} then extended this to the BYM CAR model which accounts for both spatial and non-spatial heterogeneity by assuming %This allows for greater flexibility in modeling, as it can capture variations that are not solely driven by spatial dependence, providing a more comprehensive understanding of the underlying disease distribution.
%That is,
\begin{align}
  \theta_{i} \mid \mu_{i}, z_{i}, \tau^2 \sim \N \left( \mu_i + z_{i}, \tau^2 \right),
  \label{eq:car}
\end{align}
where $\mu_{i}$ is a baseline mean  (e.g., a simple intercept parameter or potentially a regression of one or more covariates) and %{\color{blue}$\bx_i$ is a vector} of region-specific covariates with corresponding regression coefficients, $\balpha$, % denotes the associated regression coefficients governing the mean structure of the model,
%and 
$\bz \sim \CAR\left(\sig^2\right)$ % = \left(z_{1}, \ldots, z_{I} \right)^T$ represents the collection of spatial random effects and is specified as
with a density function of the form
\begin{align}
  p \left( \bz \given \sig^2 \right) \propto \left( \sig^2 \right)^{-\frac{I-1}{2}} \exp \left[ - \frac{1}{2 \sig^2} \bz^T \left( \bD - \bW \right) \bz\right],\label{eq:car_joint}
\end{align}
where $\bW = \left(w_{ij}\right)_{I \times I}$  and $\bD = \operatorname{diag}\left(\sum_{j} w_{ij}\right)$. As is customary, we let $w_{ij} = 1$ if regions $i$ and $j$ ($i \neq j$) are neighbors and $w_{ij} = 0$ otherwise. The joint distribution in~\eqref{eq:car_joint} then yields
\begin{align}
z_{i} \given \bz_{(i)}, \sig^2 &\sim \N \left( \frac{1}{m_i} \sum_{j \sim i} z_{j}, \frac{\sig^2}{m_i} \right), \label{eq:car_cond}
\end{align}
where $j \sim i$ indicates that regions $i$ and $j$ are neighbors, and $m_i = \sum_{j \sim i} w_{ij} $ denotes the number of neighboring regions for region $i$.
To quantify the informativeness of the BYM CAR model, \citet{bym:info} aimed to approximate the variance of $\theta_i$ given the remaining $\btheta_{(i)}$ by
%we first need to derive the conditional distribution of $\theta_{i}$ given the remaining $\btheta_{(i)}$ by %while averaging out the uncertainty contributed by the random effect, we integrate
integrating out $\bz$. %, allowing us to compare the informativeness of this prior with that of the gamma or the lognormal prior, which does not incorporate the spatial random effect.
%While this would not be straightforward in a general setting,
To achieve this, %\citet{bym:info}
the authors assumed a scenario in which the neighbors of region $i$ did not have any additional neighbors, thereby simplifying the necessary integration.  %this provides a lower bound on the amount of information contributed by the model's spatial structure and facilitates the derivation of a closed form expression for the variance of the conditional distribution of $\theta_{i} = \log(\lam_{i})$, which
Doing so yields a normal distribution for $\theta_i \given \btheta_{(i)}$ with $V \left[ \theta_{i} \given \btheta_{(i)}, \mu_{i}, \tau^2, \sig^2 \right] = \tau^2 + \left( \tau^2 + \sig^2 \right) / m_i$.
Plugging this into the expression in \eqref{eq:info_lognorm} implies the informativeness of the BYM CAR model can be written as:
\begin{align*}
  \widehat{a}_{i} = \frac{1}{\exp \left( V\left[ \theta_{i} \given \btheta_{(i)},  \mu_{i}, \tau^2, \sig^2 \right] \right) - 1}
  = \frac{1}{\exp \left( \tau^2 + \left( \tau^2 + \sig^2 \right) / m_i \right) - 1}.
\end{align*}
Finally, in an effort to provide a \textit{single} estimate of model informativeness---and to facilitate comparisons between analyses of different data on different spatial domains---the authors proposed substituting $m_0 = 3$ for $m_i$ as a rule-of-thumb for a baseline number of neighbors, leading to the following approximation of the \textit{overall} model informativeness:
\begin{align}
  \widehat{a}_{0} = \frac{1}{\exp \left( \tau^2 + \left( \tau^2 + \sig^2 \right) / m_0 \right) - 1} \label{eq:car_info0}.
\end{align}

In this context, recent work by \citet{quick:reliable} proposed criteria for estimates of event rates to be deemed ``reliable'' %a definition for a ``reliable'' estimate of event rates
that are based on the \textit{relative precision} (RP) measure introduced by \citet{bym:info}. %, which is connected to the model informativeness. T
Briefly, the relative precision is related to the coefficient of variation and is
defined as the ratio of the posterior median of the event rate $\lambda$ to the width of its $(1-\alpha)\times100\%$ credible interval (with $\alpha = 0.05$ by default).  Under the framework from~\eqref{eq:lam_post_gamma}, the relative precision is a monotonically increasing function of $a_{i} + y_{i}$, and \citet{bym:info} demonstrated the accuracy of the informativeness measures in~\eqref{eq:info_lognorm} and~\eqref{eq:car_info0} by comparing the relative precision of estimates under those models to the analogous relative precision curves based on~\eqref{eq:lam_post_gamma}.
%applicable across different rate models.
%They illustrated that as the model informativeness (measured by $a$) increases, the distribution narrows, leading to greater relative precision.
Based on existing reliability thresholds, \citet{quick:reliable} defined an estimate as ``reliable'' at the $(1-\alpha)$-level when its corresponding relative precision exceeds one; e.g., under the default $0.95$-level, $y_i+a_i\ge 16$ is sufficient to achieve a ``reliable'' estimate.
%median rate estimate exceeds the width of its $(1-\alpha) \times 100\%$ credible interval (i.e., RP $>1$).
%As the model informativeness increases, the relative precision increases: for example, when the model informativeness is greater than 16, the rate estimates for regions with zero events can still be considered ``reliable'' at the $\alpha = 0.05$ level.
Because these criteria could yield so-called ``reliable'' estimates when $y_i=0$ if $a_i\ge16$---illustrating the risks of overly informative models---\citet{quick:reliable} used this framework as the basis for imposing restrictions on the model's informativeness.

\subsection{Informativeness of the MCAR Model}\label{sec:mcar info}

While previous work has focused on evaluating the model informativeness in univariate settings, when applying the MCAR model to \textit{multiple} correlated groups (or outcomes), the informativeness for each group is influenced by both the spatial structure and the between-group correlations. As mentioned in Section~\ref{sec:intro}, these dependencies should result in greater informativeness in the multivariate model compared to the univariate model.

\subsubsection{Multivariate Lognormal Prior}\label{sec:lognorm approx}

Now we return to the scenario presented in the beginning of Section~\ref{sec:methods}---i.e., we have data from multiple groups, and we believe the underlying event rates for these groups are correlated (e.g., due to shared risk factors). Although our ultimate goal is to account for spatial structure and between-group dependencies simultaneously, we will first consider a multivariate, spatially independent model specification of the form:
\begin{align}
\btheta_i = \log(\blam_i) \sim \N \left(\bmu_{i}, \bPhi \right),\label{eq:mln}
\end{align}
where $\blam_{i} = \left( \lam_{i1}, \ldots, \lam_{iK} \right)^T$ represents the vector of rates for region $i$ and $\bmu_i$ and $\bPhi$ represent the length $K$ mean vector and the $K \times K$ covariance matrix on the log scale, respectively. Now suppose we are interested in calculating the informativeness of this model for group $k$, $\widehat{a}_{ik}$.  For illustration purposes, we focus on group $k=K$ and begin by deriving the conditional variance of $\theta_{ik}$ given the remaining $\theta_{ik'}$, for all $k'\ne k$, denoted by $\btheta_{i\left(k\right)}$. This can be obtained by first partitioning $\bmu_i$ and $\bPhi$ as
\begin{align}
\bmu_i =
\begin{pmatrix}
  \bmu_{i(K)} \\
  \mu_{iK}
\end{pmatrix}
\;\;\text{and}\;\;
\bPhi =
\begin{pmatrix}
  \bPhi_{(K),(K)} &  \bPhi_{(K), K} \\
   \bPhi_{K, (K)} & \phi^2_{K}
\end{pmatrix}, \label{eq:mln-mean-var}
\end{align}
which implies that $\theta_{iK} \given \btheta_{i(K)} \sim \N \left(\mu_{iK \given i(K)}, \phi^2_{K \given (K)} \right)$, where
\begin{align}
\mu_{iK \given i(K)} &= \mu_{iK} + \bPhi_{K, (K)} \bPhi_{(K),(K)}^{-1} \left(\btheta_{i(K)} - \bmu_{i(K)} \right) \notag\\
\text{and}\;\; \phi^2_{K \given (K)} &= \phi^2_{K} -  \bPhi_{K, (K)} \bPhi_{(K),(K)}^{-1} \bPhi_{(K), K}. \label{eq:mvnorm_condvar}
\end{align}
Thus, based on the approximation from~\eqref{eq:info_lognorm} and the logic of \citet{bym:info}, we can approximate the informativeness $a_{ik}$ of the model \eqref{eq:mln} for all $k$ by:
\begin{align}
  \widehat{a}_{ik} = \frac{1}{ \exp \phi^2_{k \given (k)} - 1 } = \frac{1}{\exp \left( \phi^2_{k} -  \bPhi_{k, (k)} \bPhi_{(k),(k)}^{-1} \bPhi_{(k), k} \right) - 1}.
  \label{eq:info-logmvn}
\end{align}

\subsubsection{Measuring MCAR Model Informativeness}\label{sec:mcar approx}

Having established a framework for a general multivariate normal model specification for $\btheta_{i}$, we now shift our focus to the scenario:
\begin{align}
  \btheta_{i\cdot} \sim \N\left(\bmu_i + \bz_{i\cdot},\, \bPsi \right), \label{eq:mcar}
\end{align}
where %{\color{orange}$\bmu_i$ is a baseline mean vector}, 
$\bPsi$ is a $K\times K$ diagonal matrix with elements $\tau_k^2$ and $\bz\sim \MCAR\left(\bSig\right)$, based on the approach of \citet{gelfand:Proper2003}, where $\bz = \left(\bz_{1 \cdot}^T, \ldots, \bz_{I \cdot}^T \right)^T$ and $\bSig$ represents a between-group covariance matrix. This implies a joint distribution of the form %. This implies a joint distribution of the form
\begin{align}
    p(\bz \given \bSig) \propto \vert \bSig \vert^{-\frac{I - 1 }{2}} \exp \left[ -\frac{1}{2} \bz^T \left[ \left( \bD - \bW \right) \otimes \bSig^{-1} \right] \bz \right], \label{eq:mcar_joint}
\end{align}
and a conditional distribution of the form
\begin{align}
\bz_{i \cdot} \given \bz_{(i) \cdot}, \bSig \sim \N \left(\frac{1}{m_i} \sum_{j \sim i} \bz_{j \cdot}, \frac{1}{m_i} \bSig \right).\label{eq:mcar_cond}
\end{align}

To quantify the information contributed by the MCAR model to region $i$ for group $k=K$, we need to estimate the variance, $V\left[\theta_{iK} \mid \btheta_{(iK)}, \bmu, \bSig, \bPsi \right]$, by integrating out $\bz$. To obtain the conditional distribution of $\theta_{iK}$, we first derive the multivariate conditional distribution of $\btheta_{i\cdot}$. Following \citet{bym:info}, we consider the case where $m_j = 1$ for all $j \sim i$ in order to obtain the precision. As demonstrated in Appendix~A, this yields
\begin{align}
    \btheta_{i \cdot} \given \btheta_{(i) \cdot}, \bmu, \bPsi, \bSig \sim \N \left( \bmu_{i \cdot} + \sum_{j \sim i} \frac{1}{m_i} \left( \btheta_{j \cdot} - \bmu_{j \cdot} \right), \bPsi +  \frac{1}{m_{i}} \left( \bPsi + \bSig \right) \right) \label{eq:full_cond1}.
\end{align}
Next, we define $\bPhi_{i} = \bPsi +  \frac{1}{m_{i}} \left( \bPsi + \bSig \right)$ and partition the mean vector and covariance matrix from~\eqref{eq:full_cond1} as follows:
\begin{align*}
  \btheta_{i \cdot} & \given \btheta_{(i) \cdot}, \bmu, \bPhi_{i}  \sim \N
  \left(
    \begin{pmatrix}
      \bmu_{i(K)} + \sum_{j \sim i} \frac{1}{m_i} \left( \btheta_{j(K)} - \bmu_{j(K)} \right) \\
      \mu_{iK} + \sum_{j \sim i} \frac{1}{m_i} \left( \mbox{$\theta$}_{jK} - \mbox{$\mu$}_{jK} \right)
    \end{pmatrix}
    ,
    \begin{pmatrix}
      \bPhi_{i(K), i(K)} & \bPhi_{i(K), iK} \\
       \bPhi_{iK, i(K)}&  \phi^2_{iK}
    \end{pmatrix}
  \right).
\end{align*}
Following the result from~\eqref{eq:mvnorm_condvar}, the conditional variance of $\theta_{iK}$ given $\btheta_{(iK)}$, denoted $\phi^2_{iK \given (iK)}$, can be expressed as
\begin{align}
%    V[ \theta_{iK} & \given \btheta_{(iK)}, \bmu, \bPhi ]
%         =
\phi^2_{iK \given (iK)} &=  \phi^2_{iK} - \bPhi_{iK, i(K)}\bPhi_{i(K), i(K)}^{-1}\bPhi_{i(K), iK} \notag\\ %\label{eq:condvar}\\
        &= \tau_{K}^2 + \frac{1}{m_{i}} \left\{\tau^2_{K} + \sig^2_{K} - \bSig_{K, (K)} \left[ (m_{i} + 1) \bPsi_{(K, K)} + \bSig_{(K, K)} \right]^{-1} \bSig_{(K), K} \right\}  \label{eq:condvar-full}.
    \end{align}
% This expression accounts for the dependency structures of different outcomes in the spatial effect $\bz$ through the covariance matrix $\bSig$.
We can see that as the number of neighbors ($m_i$) increases and/or when the dependencies among groups (as measured by $\bSig_{(K), K}$) are stronger, the conditional variance decreases. Now using this conditional variance, the model informativeness of region $i$ for $K^{th}$ group is given by:
\begin{align}
    \widehat{a}_{iK}
    & = \frac{1}{\exp \phi^2_{iK \given (iK)} - 1} \nonumber \\
    & = \frac{1}{ \exp \left[ \tau_{K}^2 + \frac{1}{m_{i}} \left\{\tau^2_{K} + \sig^2_{K} - \bSig_{K, (K)} \left[ (m_{i} + 1) \bPsi_{(K, K)} + \bSig_{(K, K)} \right]^{-1} \bSig_{(K), K} \right\} \right] - 1
    }. \label{eq:info-full}
\end{align}
To quantify the overall informativeness of the model and enable comparisons across analyses involving different datasets and spatial domains, we replace $m_i$ with the ``rule-of-thumb'' number of neighbors, $m_0 = 3$, as in \citet{bym:info}. This would imply that now instead of $\bPhi_{i} = \bPsi +  \frac{1}{m_{i}} \left( \bPsi + \bSig \right)$, we work with $\bPhi = \bPsi + \frac{1}{m_{0}} \left( \bPsi + \bSig \right)$. Additionally, replacing $K$ with $k$ to denote the expression for any group, we obtain:
\begin{align}
    \widehat{a}_{0k}
    &= \frac{1}{ \exp \left[ \tau_{k}^2 + \frac{1}{m_{0}} \left\{\tau^2_{k} + \sig^2_{k} - \bSig_{k, (k)} \left[ (m_{0} + 1) \bPsi_{(k, k)} + \bSig_{(k, k)} \right]^{-1} \bSig_{(k), k} \right\} \right] - 1
    }. \label{eq:info0-full}
\end{align}
It can then be seen that the \textit{univariate} BYM  CAR model in \eqref{eq:car_info0} is a special case of \eqref{eq:info0-full} when the groups are independent (i.e., $\bSig_{(k), k} = \bzero$).

\section{Restricting MCAR Model Informativeness}\label{sec:res-mcar}

While we want to borrow strength across groups and account for spatial autocorrelation to model the rates, a key objective of this work is to guard against oversmoothing. As implied by~\eqref{eq:info0-full}, however, any restriction on $\widehat{a}_{0k}$ will require restrictions on both the spatial covariance matrix, $\bSig$, and the non-spatial variance parameters, $\tau_k^2$ (via $\bPsi$). With that in mind, we will first demonstrate how to efficiently impose restrictions on the general multivariate lognormal setup discussed in Section~\ref{sec:lognorm approx} before presenting our approach for implementing a restricted MCAR model.

\subsection{Implementing Restrictions for Multivariate Normal Models}\label{sec:restrict_mlm}

Suppose we model the log event rates using the framework from~\eqref{eq:mln} where $\btheta_{i\cdot}\sim \N\left(\bmu_i,\bPhi\right)$ and we wish to fit our model under a restriction in which $\widehat{a}_{0k} < A_k$, for $k=1,\ldots,K$. Based on the expression for $\widehat{a}_{0k}$ in~\eqref{eq:info-logmvn}, this will require sampling the $K\times K$ covariance matrix $\bPhi$ while simultaneously satisfying $K$ sets of criteria.  This poses challenges because a matrix-wide accept/reject strategy (e.g., in which candidate matrices are sampled from an inverse Wishart distribution and accepted if they satisfy the necessary criteria) is likely to be inefficient with high rejection rates---particularly when the thresholds, $A_k$, are small---and an element-by-element strategy (i.e., sampling the variance and correlation parameters directly) would lack conjugate priors and require care to ensure positive definiteness.

\begin{comment}
%Assume we are modeling the rates as defined in equation~\eqref{eq:mln}. To achieve restriction, we aim to ensure that, $\widehat{a}_{ik}$ in \eqref{eq:info-logmvn}, remains below a specified threshold, $A_{k}$. However, sampling covariance matrices that satisfy these constraints is challenging. For instance, while an inverse Wishart prior would allow us to sample a covariance matrix directly from its full conditional distribution, this approach won't respect constraints on the model informativeness. As such, ensuring that our covariance matrix, $\bPhi$, satisfies $\widehat{a}_{0k} < A_k$ for all $k$ may require an inefficient accept/reject sampling process.

%Alternatively, we could sample each element in the covariance matrix separately---i.e., sample the individual variances and the correlations. However, this approach has two main drawbacks. First, because many of these parameters lack conjugate priors, it necessitates inefficient Metropolis-Hastings updates. Second, we must ensure that sampled $\bSig$ matrices remain positive-definite. Second, we must ensure that the sampled $\bSig$ matrices remain positive definite. For example, a covariance matrix is invalid if $\sig_i^2 \sig_j^2 < \sig_i^2 \sig_j^2 \rho_{ij}^2$ for any pair $i \ne j$ \citep{rue:Gaussian2005}. Ensuring that this condition holds simultaneously for all pairs can be challenging when sampling from a truncated distribution designed to satisfy the informativeness restriction, since informativeness is a function of the conditional variance rather than the marginal variance.
\end{comment}

Thus, we implement an approach that partitions the joint distribution of $\btheta_i$ into a series of marginal and sequentially conditional distributions that we refer to as the \emph{regression approach}.  Specifically, it can be shown via properties of the multivariate normal distribution that the joint distribution for $\btheta_{i\cdot}$ in~\eqref{eq:mln} can be rewritten as:
\begin{align*}
  \theta_{i1} & \sim  \N \left( \mu_{i1}, \phi^2_{1} \right) \\
  \theta_{i2} \given \theta_{i1} & \sim  \N \left( \mu_{i2} + \gamma_{1, 2} \left( \theta_{i1} - \mu_{i1} \right), \phi^2_{2 \given 1} \right) \\
  \theta_{i3} \given \theta_{i1}, \theta_{i2} & \sim \N \left( \mu_{i3} + \gamma_{1, 3} \left( \theta_{i1} - \mu_{i1} \right) + \gamma_{2, 3} \left( \theta_{i2} - \mu_{i2} \right), \phi^2_{3 \given 1,2} \right) \\
  & \vdots \\
  \theta_{iK} \given \btheta_{i,<K} & \sim  \N \left( \mu_{iK} + \bgamma_{K}^{T} \left( \btheta_{i,<K} - \bmu_{i,<K} \right), \phi^2_{K \given <K} \right),
\end{align*}
where $\btheta_{i,<k} = \left( \theta_{i,1}, \ldots, \theta_{i,k-1} \right)^{T}$ and $\bmu_{i,<k} = \left(\mu_{i,1}, \ldots, \mu_{i,k-1} \right)^{T}$ for $k>1$. Moreover, we define $\bgamma_k = \left(\gamma_{1, k}, \gamma_{2,k}, \ldots, \gamma_{k-1, k} \right)^T = \left[ \bPhi_{<k,<k} \right]^{-1}\bPhi_{<k,k}$ and 
\begin{align}
\phi^2_{k \given <k} &= \phi^2_k - \bPhi_{k,<k}\left[\bPhi_{<k,<k}\right]^{-1}\bPhi_{<k,k},
\label{eq:cond_var_mlm}
\end{align}
where $\bPhi_{<k,k}$ denotes the vector formed by the first $(k-1)$ entries of the $k$th column of $\bPhi$, and $\bPhi_{<k,<k}$ denotes the $(k-1)\times(k-1)$ leading principal submatrix of $\bPhi$.
% where for $k>1$ $\btheta_{i, <k} = \left( \theta_{i,1}, \ldots, \theta_{i, k-1} \right)^{T}$, $\bmu_{i,<k} = \left(\mu_{i, 1}, \ldots, \mu_{i, k-1} \right)^{T}$, $\bbeta_{k}  = \left[ \bPhi_{<k, <k} \right]^{-1} \bPhi_{<k, k}$, and
% \begin{align}
%   \phi^2_{k \given <k} &= \phi^2_{k} -  \bPhi_{k, <k} \left[ \bPhi_{<k, <k} \right]^{-1} \bPhi_{<k, k}
%   \label{eq:cond_var_mlm},
% \end{align}
% where $\bPhi_{<k, k}$ is a vector containing the first $(k-1)$ rows of the $k$th column of $\bPhi$, and $\bPhi_{<k, <k}$ is the submatrix of $\bPhi$ consisting of the first $(k-1)$ rows and columns.  
Finally, it should be noted that conjugate priors are available for the elements of $\bgam_{k}$ and the $\phi^2_{k \given < k}$, for $k = 2, \ldots, K$---in addition to the elements of $\bmu$ and $\phi_1^2$---and no additional conditions are needed on these parameters to ensure $\bPhi$ is positive definite. % (e.g., the decomposed inverse Wishart formulation)s

It should also be noted that this reparameterization is not itself novel.  For instance, \citet{eaves:Priors1992} investigated priors for conditional variances, and \citet{jin:Generalized2005} examined a regression formulation for a bivariate spatial model. Moreover, \citet{lindley:Bayesian1978} discussed the regression interpretation of the multivariate normal distribution, which was later reiterated by \citet{brown:inference1994} through the use of the inverse Wishart and \textit{generalized} inverse Wishart priors that accommodate the regression framework.
%{\color{blue}That said, the novelty of this framework}
As such, the novelty of this work is how we can use the regression approach to implement restrictions of the form $\widehat{a}_{0k} < A_{k}$.  To do so, we begin by rearranging the terms in~\eqref{eq:info-logmvn} as
%Now, we show how we can restrict the parameters so that we ensure the informativeness does not exceed a threshold. Again, this suggests the equation~\eqref{eq:info-logmvn} is less than $A_{k}$, which implies
\begin{align}
  \phi^2_{k \given (k)} > \log(1 \slash A_{k} + 1) = A_{\phi, k}
  \label{eq:rule_lml}.
\end{align}
That said, note the restriction rule in \eqref{eq:rule_lml} is stated in terms of $\phi^2_{k \given (k)}$, the \textit{full} conditional variance, not $\phi^2_{k \given < k}$, the \textit{partial} conditional variance identified in \eqref{eq:cond_var_mlm}. Therefore, we must first find an expression for $\phi^2_{k \given (k)}$ written as a function of $\phi^2_{k \given <k}$ and $\bgam_{k}$ before we can impose the restriction rule for $\phi^2_{k \given < k}$ using \eqref{eq:rule_lml}. To do so, we note that the conditional distribution we desire can be expressed as:
\begin{align*}
    p \left( \theta_{ik} \given \btheta_{i(k)},\bmu,\bPhi \right) \propto & p\left( \theta_{ik} \given \btheta_{i < k},\bmu_{i,<k},\bgamma_{k},\phi_{k\given <k}^2 \right)\\
    &\times \prod_{k' > k}^{K} p \left( \theta_{ik'} \given \mu_{ik'} + \bgamma_{k'}^{T} \left(\btheta_{i, < k'}  - \bmu_{i,< k'} \right), \phi^2_{k' \given < k'}\right),
\end{align*}
which yields a normal distribution with variance
  \begin{align}
    V \left[ \theta_{ik} \given \btheta_{i(k)},\bmu, \bPhi \right] = \phi^2_{k \given (k)} = \left[ \frac{1}{\phi^2_{k \given < k}} + \sum_{k' > k}^{K}\frac{\gamma^2_{k, k'}}{\phi^2_{k' \given < k'}} \right]^{-1}.
    \label{eq:condvar_reg_theta}
  \end{align}
From this, it can be shown that the restriction rule in \eqref{eq:rule_lml} requires
  \begin{align}
    \phi^2_{k \given < k} > \left[\frac{1}{A_{\phi, k}} -  \sum_{k' > k}^{K} \frac{\gamma^2_{k, k'}}{\phi^2_{k' \given < k'}}\right]^{-1} = A^{*}_{\phi_k, k}.
    \label{eq:rule_lml_phi_k}
  \end{align}
Unfortunately, because $\phi^2_{k \given < k}$ also appears in the expression $V \left[ \theta_{i \ell} \given \btheta_{i(\ell)}, \bmu,\bPhi \right]$, where $\ell < k$, we also need to restrict $\phi^2_{k \given <k}$ such that
  \begin{align}
    V \left[ \theta_{i \ell} \given \btheta_{i (\ell)},\bmu, \bPhi \right] 
    = \left[ \frac{1}{\phi^2_{\ell \given < \ell}} + \sum_{k' > \ell}^{K}\frac{\gamma^2_{\ell, k'}}{\phi^2_{k' \given < k'}} \right]^{-1} > A_{\phi, \ell} \nonumber \\
    \Rightarrow \phi^2_{k \given < k} > \left[\frac{1}{A_{\phi, k}} -  \frac{1}{\phi^2_{\ell \given < \ell}} - \sum_{\substack{k' > \ell \\ k' \ne \ell}}^{K}\frac{\gamma^2_{\ell, k'}}{\phi^2_{k' \given < k'}} \right]^{-1} \gamma_{\ell,k}^2 = A^{*}_{\phi_k, \ell}
    \label{eq:rule_lml_phi_l}
  \end{align}
Therefore, we need to ensure $\phi^2_{k \given < k} > \textrm{max} \left(A^{*}_{\phi_k, 1}, \ldots, A^{*}_{\phi_k, k-1}, A^{*}_{\phi_k, k} \right) = \tilde{A}_{\phi, k}$. Finally, note that because the inverse gamma distribution is the conjugate prior for $\phi^2_{k \given < k}$, it is straightforward to sample $\phi^2_{k \given < k}$ from a truncated version of its full-conditional distribution. %truncated to
  %satisfy the restriction rule in \eqref{eq:rule_lml}.

Identifying criteria for $\gamma_{\ell,k}$ is less complicated, as each $\gamma_{\ell,k}$ only appears in \eqref{eq:rule_lml_phi_l} for its corresponding $A_{\phi_k, \ell}^{*}$. Specifically, we aim to ensure for $k = 2, \ldots, K$ and for all $\ell < k$:
\begin{align}
  \phi^2_{k \given < k} & > \left[\frac{1}{A_{\phi, \ell}} -  \frac{1}{\phi^2_{\ell \given < \ell}} - \sum_{k' > \ell}^{K}\frac{\gamma^2_{\ell, k'}}{\phi^2_{k' \given < k'}} \right]^{-1} \gamma_{\ell,k}^2 \nonumber \\
  & \Rightarrow 0 < \gamma_{\ell,k}^2 < \phi^2_{k \given < k} \left[\frac{1}{A_{\phi, \ell}} -  \frac{1}{\phi^2_{\ell \given < \ell}} - \sum_{k' > \ell}^{K}\frac{\gamma^2_{\ell, k'}}{\phi^2_{k' \given < k'}} \right] = A_{\gamma, \ell, k} \\
  & \Rightarrow - \sqrt{A_{\gamma, \ell, k}} < \gamma_{\ell,k} <  \sqrt{A_{\gamma, \ell, k}}.
  \label{eq:rule_lml_gamms}
\end{align}
Here again, because the normal distribution is the conjugate prior for $\gamma_{\ell, k}$, sampling the $\gamma_{\ell, k}$ parameters under the restriction in~\eqref{eq:rule_lml_gamms} is straightforward.

\subsection{Implementing Restrictions for the MCAR Model} \label{sec:imp-res-mcar}

Now, we extend our approach from Section~\ref{sec:restrict_mlm} %the multivariate normal model
to the MCAR model from~\eqref{eq:mcar}. % which we ultimately aim to restrict.
To rewrite the MCAR structure using the regression approach, we first permute the elements of $\bz$ to block by groups instead of areal units: i.e., $\widetilde{\bz} = \left(\bz_{\cdot 1}^T, \ldots, \bz_{\cdot K}^T \right)^T$. Then \eqref{eq:mcar_joint} becomes:
\begin{align}
    p(\widetilde{\bz} \given \bSig)
    & \propto \vert \bSig \vert^{-\frac{I - 1 }{2}} \exp \left[ -\frac{\widetilde{\bz}^T \left[  \bSig^{-1} \otimes \left( \bD - \bW \right)  \right] \widetilde{\bz}}{2}\right]
    \label{eq:mcar_joint_2}.
\end{align}
Using the regression approach here again, we get
\begin{align}
    p (\bz_{\cdot 1}  \given \bSig)  &\propto \left(\sig^2_{1} \right)^{-\frac{I-1}{2}} \exp \left[ -  \frac{  \bz_{\cdot 1}^T \left( \bD - \bW \right) \bz_{\cdot 1}}{2 \sig^2_{1}} \right] \nonumber \\
    p (\bz_{\cdot 2}  \given \bz_{\cdot 1}, \bSig) & \propto \left( \sig^2_{2 \given 1} \right)^{-\frac{I-1}{2}}\exp \left[ -  \frac{\left( \bz_{\cdot 2} -  \beta_{1, 2} \bz_{\cdot 1} \right)^T \left( \bD - \bW \right) \left( \bz_{\cdot k} -  \beta_{1, 2} \bz_{\cdot 1}\right)}{ 2 \sig^2_{2 \given < 1}} \right] \nonumber \\
    & \vdots  \nonumber \\
    p (\bz_{\cdot k}  \given \bz_{\cdot < k}, \bSig) & \propto \left( \sig^2_{k \given <k} \right)^{-\frac{I-1}{2}}\exp \left[ -  \frac{\left( \bz_{\cdot k} -  \bz_{\cdot < k} \bbeta_{k} \right)^T \left( \bD - \bW \right) \left( \bz_{\cdot k} -  \bz_{\cdot < k} \bbeta_{k} \right)}{ 2 \sig^2_{k \given < k}} \right] \label{eq:mcar_cond1} \\
    & \vdots \nonumber \\
    p (\bz_{\cdot K}  \given \bz_{\cdot < K}, \bSig) & \propto \left( \sig^2_{K \given <K} \right)^{-\frac{I-1}{2}}\exp \left[-\ \frac{\left( \bz_{\cdot K} -  \bz_{\cdot < K} \bbeta_{K} \right)^T \left( \bD - \bW \right) \left( \bz_{\cdot K} -  \bz_{\cdot < K} \bbeta_{K} \right)}{ 2 \sig^2_{K \given < K}} \right] \nonumber,
\end{align}
where $\bz_{\cdot < k}
= [
\bz_{\cdot 1} \; \bz_{\cdot 2} \; \cdots \; \bz_{\cdot k-1}
]
\in \mathbb{R}^{I \times (k-1)}$ and $\bbeta_{k} = \left[ \bSig_{< k, < k} \right]^{-1} \bSig_{<k, k} $. Details of this derivation can be found in Appendix~B along with the full-conditionals of the parameters in the MCAR model for use in our Markov chain Monte Carlo (MCMC) algorithm.

To restrict the informativeness in the MCAR model, we need to determine the range of $\sig^2_{1}$ and each of the $\sig^2_{k \given < k}$, $\bbeta_{k}$, and $\tau^2_{k}$ parameters for $k = 2, \ldots, K$ that will ensure $\phi_{k\given (k)}^2 > \log \left( 1 \slash A_{k} + 1 \right)$,
%{\color{orange}$\widehat{a}_{0k} < A_k$},
%$> \log \left( 1 \slash A_{k} + 1 \right)$,
where $\phi_{k\given (k)}^2$ is as defined in~\eqref{eq:condvar-full}, thus satisfying $\widehat{a}_{0k}<A_k$ for all $k$.
%$\widehat{a}_{0k}$ is defined in \eqref{eq:info0-full}, thus satisfying \eqref{eq:condvar}.
The challenge here is that we do not have a consistent covariance matrix $\bPhi_{i}$ for all $i$. To overcome this, we let $\bPhi = \bPsi + \frac{1}{m_0} \left( \bPsi + \bSig \right)$ and define pseudo-regression coefficients, $\bgam_{k}$, based on $\bPhi$ such that $
  \bgamma_{k} = \left[ \bPhi_{< k, < k} \right]^{-1} \bPhi_{<k, k}  = \frac{1}{m_0} \left[ \bPhi_{< k, < k }\right]^{-1} \bSig_{<k, k}
$
since $\bPsi_{k, k'} = 0$ for all $k \ne k'$. The range for $\sig^2_{k \given < k}$, $\tau^2_{k}$, and $\bgam_{k}$ can then be determined from the range of $\phi^2_{k \given < k}$. Again, we can use truncated conjugate priors to facilitate sampling from the full-conditional distributions in a manner that ensures $\widehat{a}_{0k}<A_k$. Details on how the restriction is implemented in our MCMC algorithm are provided in Appendix~B.4.

\section{Ischemic Heart Disease Mortality in the US} \label{sec:analysis}

\begin{table}[t]
  \centering
  \begin{tabular}{l|rrr|rr}
    \hline
   & Population & Deaths & Rate (per 100,000) & \% $y_{ik}< 10$ & \% $y_{ik}=0$ \\
%   & Population & Deaths & Crude Rate\footnotemark[1] & $<$ 10 Deaths\footnotemark[2] & Zero Deaths\footnotemark[3] \\
   \hline
   White Males & 9,126,899 & 53,472 & 586 & 67\% & 7\% \\
   White Females & 10,303,383 & 18,345 & 178 & 88\% & 26\% \\
   Black Males & 854,366 & 5,021 & 588 & 97\% & 71\% \\
   Black Females & 1,060,137 & 3,346 & 316 & 98\% & 77\% \\
   \hline
  \end{tabular}
  \caption{Summary of ischemic heart disease mortality by race and sex in (contiguous) US counties for adults aged 55--64 in 1980.}
  \label{tab:hd_desc_table}
  %\caption*{\parbox{\textwidth}{%
  %\small{\footnotemark[1] (Deaths) $\slash$ (Population) $\times 100\small{,}000$.\\
  %\footnotemark[2] the percentage of counties reporting fewer than 10 deaths. \\
  %\footnotemark[3] the percentage of counties reporting no deaths.}
  %}}
  \end{table}

To illustrate the impact and the evaluation of the model informativeness on rate estimates, we consider a dataset comprised of county-level ischemic heart disease-related death counts among Black and White men and women ($K = 4$) aged 55--64 from the $I = \text{3,109}$ counties of the contiguous United States in 1980.  The data are available in the US Centers for Disease Control and Prevention (CDC) WONDER database at \url{https://wonder.cdc.gov/mortsql.html}. These data were derived from publicly available mortality data resources maintained by the CDC. While previous studies have demonstrated the existence of geographic disparities in the rate of death due to \emph{all forms} of heart disease by race and sex \citep{quick:waller,adam:arg} and by age \citep{adam:arg} or compared trends in \emph{specific} forms of heart disease \citep[e.g.,][]{roth:cvd}, few studies have considered the analysis of a specific form of heart disease (ischemic hearth disease; ICD-9 410--414) for a single ten-year age group (individuals 55--64), stratified by race (Black and White) and sex (female and male) due to the sparsity of such data. Moreover, analyzing county-level data from across the contiguous US, stratified by race and sex, captures substantial heterogeneity in population size and the number of deaths per county. This allows for a deeper exploration of the MCAR model’s behavior while enabling groups to borrow strength across subpopulations and assessing the reliability, as defined in \citet{quick:reliable}, in scenarios ranging from counties with high rates and large populations (White men) to those with low rates and small populations (Black women). For example, as shown in Table~\ref{tab:hd_desc_table}, 98\% of counties reported fewer than 10 deaths due to ischemic heart disease among Black women compare to 67\% for White men. Finally, while recent vital statistics data would certainly be of \emph{epidemiologic} interest, the CDC imposes privacy protections on vital statistics data that date back to 1989 in the form of the suppression of small counts \citep{cdc:sharing}. Because analyzing data with small counts is the motivation for this work, analyzing data that are not subject to these privacy protections allows us to freely discuss specifics of the data as the data themselves are publicly available. Thus, we believe an analysis of these data will provide a compelling illustration of the dangers of analyzing such data using the CAR and MCAR models and the benefits of restricting the model's informativeness in conjunction with the reliability definition provided in \citet{quick:reliable}.
\begin{comment}we believe this represents the first analysis of geographic trends in a specific form of heart disease (ischemic heart disease; ICD-9 410--414) for a single ten-year age group (individuals 55--64), stratified by race (Black and White) and sex (female and male). \end{comment} 
%\subsection{Analysis Plan}\label{sec:analysis}

We now demonstrate how the MCAR model estimates county-level mortality rates by borrowing information across the four subgroups, and compare its estimates to those from the CAR model. Additionally, we illustrate how the MCAR model tends to yield higher levels of informativeness due to its dependency structure, necessitating control over model informativeness to prevent overly smooth estimates while still leveraging the benefits of the multivariate model. To address this, we fit both standard and restricted versions of the BYM CAR and MCAR models. To focus on the smoothing behavior of the random effects under different levels of model informativeness, we assume an intercept-only model for the baseline mean $\mu_i$ in \eqref{eq:car}---i.e., $y_{ik} \sim \Pois \left( n_{ik} \lam_{ik}\right)$, where $\theta_{ik} = \log \lam_{ik} \sim \N \left( \alpha_{0k} + z_{ik}, \tau^2_{k}\right)$, for $i = 1, \ldots, I$ and $k = 1, \ldots, K$. For the CAR model, we assume that $\bz_{\cdot k} \sim \CAR \left( \sig^2_{k} \right)$ for each $k$, and $\bz \sim \MCAR \left(\bSig \right)$ for the MCAR model.

For the priors, we use a flat prior for $\alpha_{0k}$ and vague priors consistent with \citet{bernardinelli} and \citet{waller:Hierarchical1997} for the variance parameters. Specifically, for the BYM CAR model, we assume $\tau^2_{k} \sim \IG \left(1, 1 \slash 100 \right)$ and $\sig^2_{k} \sim \IG \left( 1, 1 \slash 7 \right)$. For the MCAR model, we assume $\bSig \sim \IW \left( \nu, \bG \right)$, where $\nu = K + 1$ and $\bG$ is a $K \times K$ diagonal matrix with diagonal entries set to $2 \slash 7$, reflecting the insufficient information to specify the dependency. The use of the inverse Wishart prior on $\bSig$ should avoid the ``ordering'' issue described in \citet{jin:Generalized2005}, and the choice of hyperparameters ensures that the prior for $\bSig$ under the MCAR model is comparable to the priors used for $\sig_k^2$ in the CAR model. We fit our models using a MCMC algorithm run for
%Our MCMC algorithm runs for
100,000 iterations, discarding the first half as burn-in and saving every 5th iteration to reduce memory usage. We first fit the standard model without imposing any restrictions to evaluate the model's informativeness. We then refit the model under the restricted framework, applying the constraints: $\widehat{a}_{0k} < 6, 12, 18, \ldots, 42, 48$ for $k = 1, \ldots, K$. The most restrictive constraint, $\widehat{a}_{0k} < 6$, is based on the reliability criteria of \citet{quick:reliable}, and the restrictions were selected to provide a continuum with regards to the model's informativeness for comparison purposes.

%  We also examine whether the ordering of the regression approach impacts the results by comparing two specific orders. In the first order, we model Black females, Black males, White females, and White males, naming this sequence ``increasing'' as the overall death count generally increases in this order. For comparison, we reverse this sequence and label it ``decreasing.''

%\subsection{Results}\label{sec:results}

\begin{figure}[t]
  \begin{center}
      \includegraphics[width=.75\textwidth]{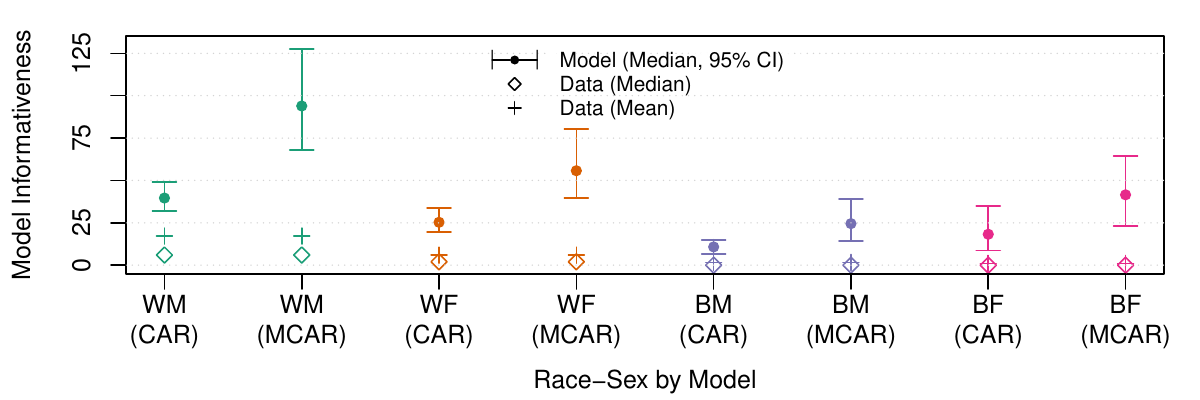}
  \end{center}
  \caption{Comparison of the estimated CAR and MCAR model informativeness.}
  \label{fig:info_ci}
\end{figure}

Figure~\ref{fig:info_ci} summarizes and compares the estimated model informativeness for both the CAR and MCAR model across the subgroups in our study.  Here, a few key findings jump out.  First, the estimated model informativeness, as measured by $\widehat{a}_{0k}$, is larger under the MCAR model than under the CAR model for all groups.  As discussed in Section~\ref{sec:mcar approx}, this is to be expected, as accounting for the between-group dependencies in the model should \emph{reduce} the conditional variance of the log death rate parameters, $\theta_{ik}$, thus \emph{increasing} the informativeness of the model.  Second, we see that both the CAR and MCAR modeling frameworks often yield models which overpower the data in most counties, as the estimated model informativeness exceeds the median and mean death count for all groups.  This suggests that oversmoothing---regardless of how it is defined---is likely to be of concern in the estimates produced by these models.  Finally, because $\widehat{a}_{0k} > 16$ for all groups from both models except for Black males using the CAR model, we may obtain ``reliable'' estimates based on the criteria of \citet{quick:reliable}, when the observed counts, $y_{ik}$, are small or zero, reinforcing our concerns about oversmoothing.

\begin{figure}[t]
  \begin{center}
      \subfigure[MCAR; White Males]{\includegraphics[width=.45\textwidth]{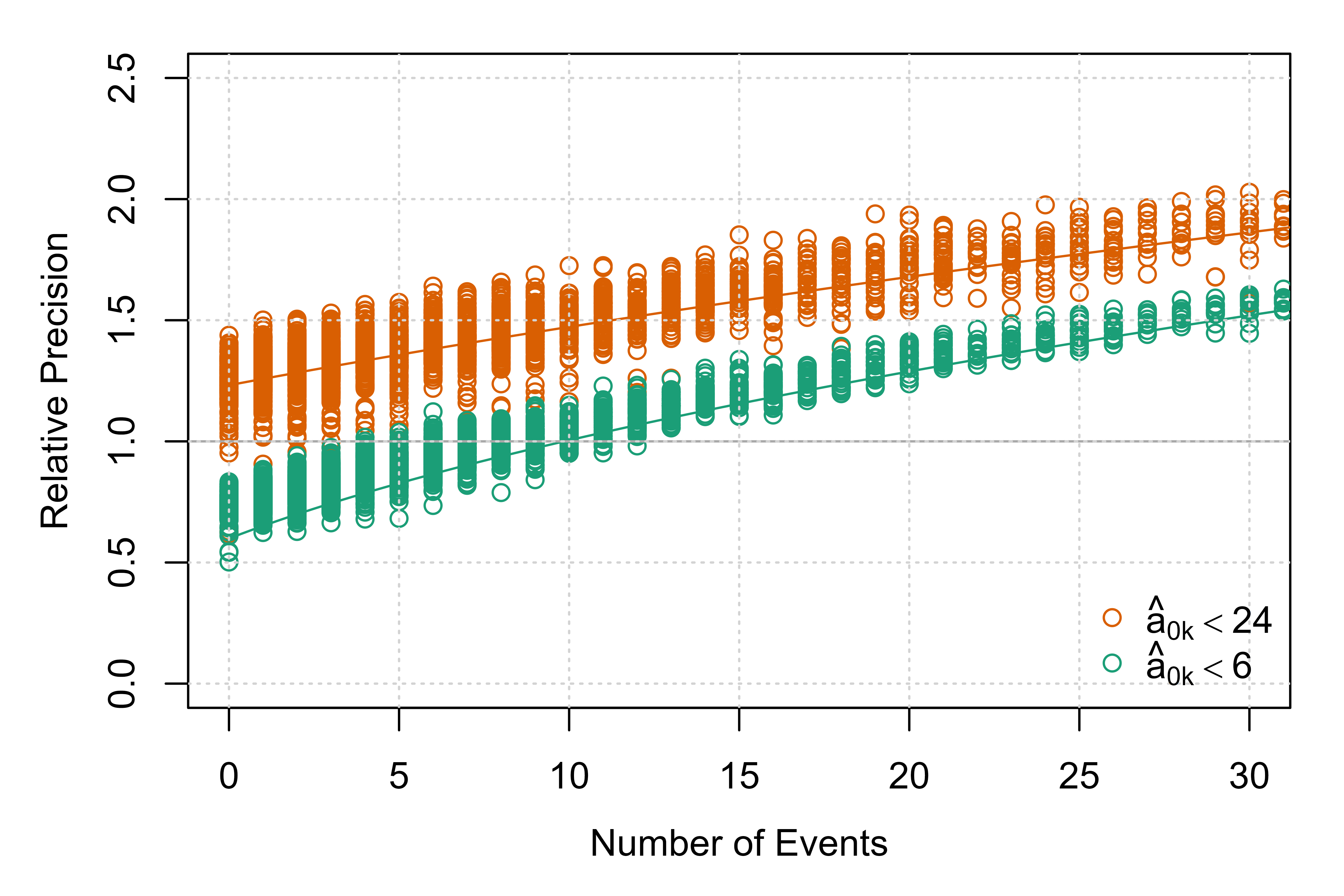}\label{fig:rp_MCAR_wm}}
      \subfigure[MCAR; White Females]{\includegraphics[width=.45\textwidth]{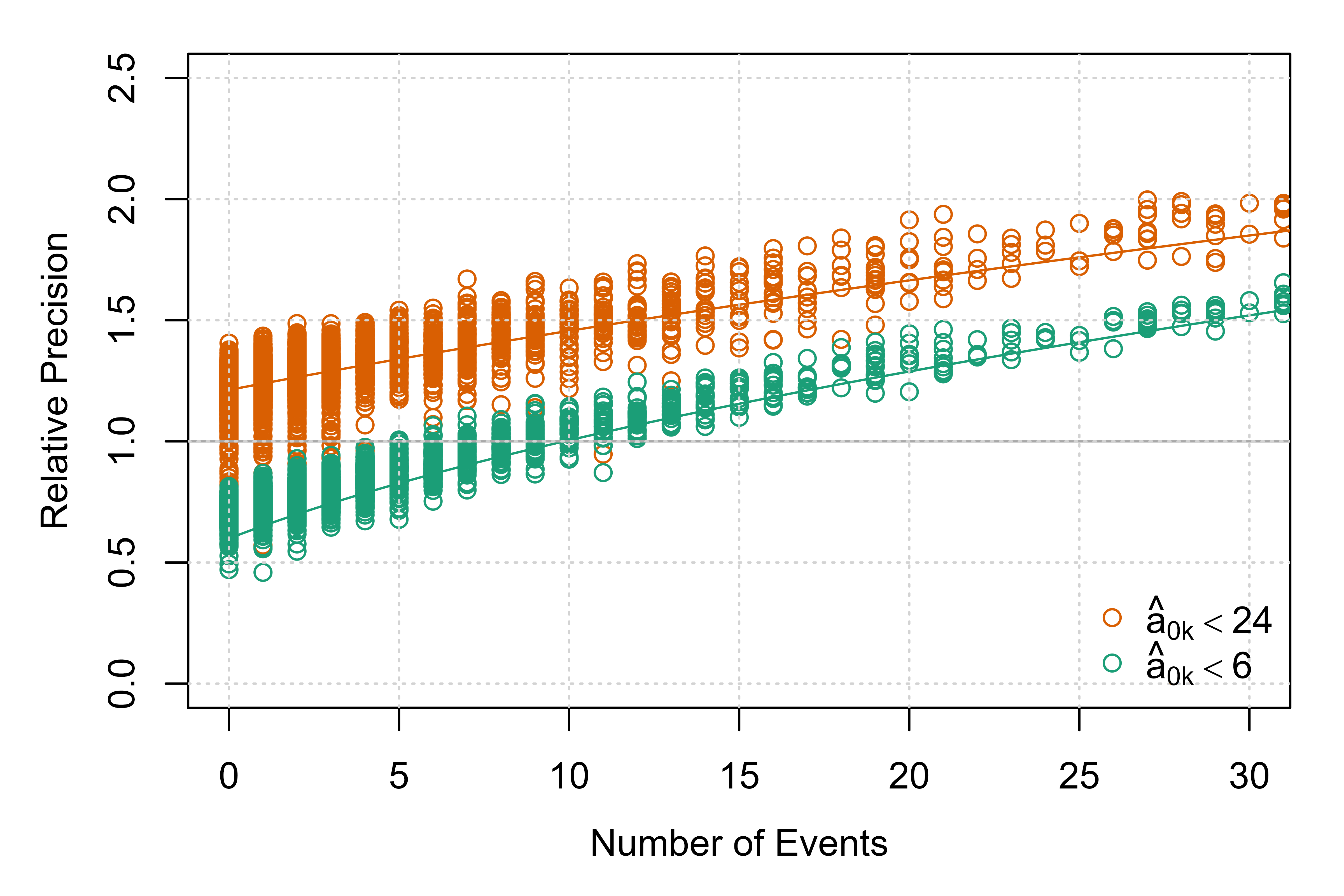}\label{fig:rp_MCAR_wf}}
      \subfigure[MCAR; Black Males]{\includegraphics[width=.45\textwidth]{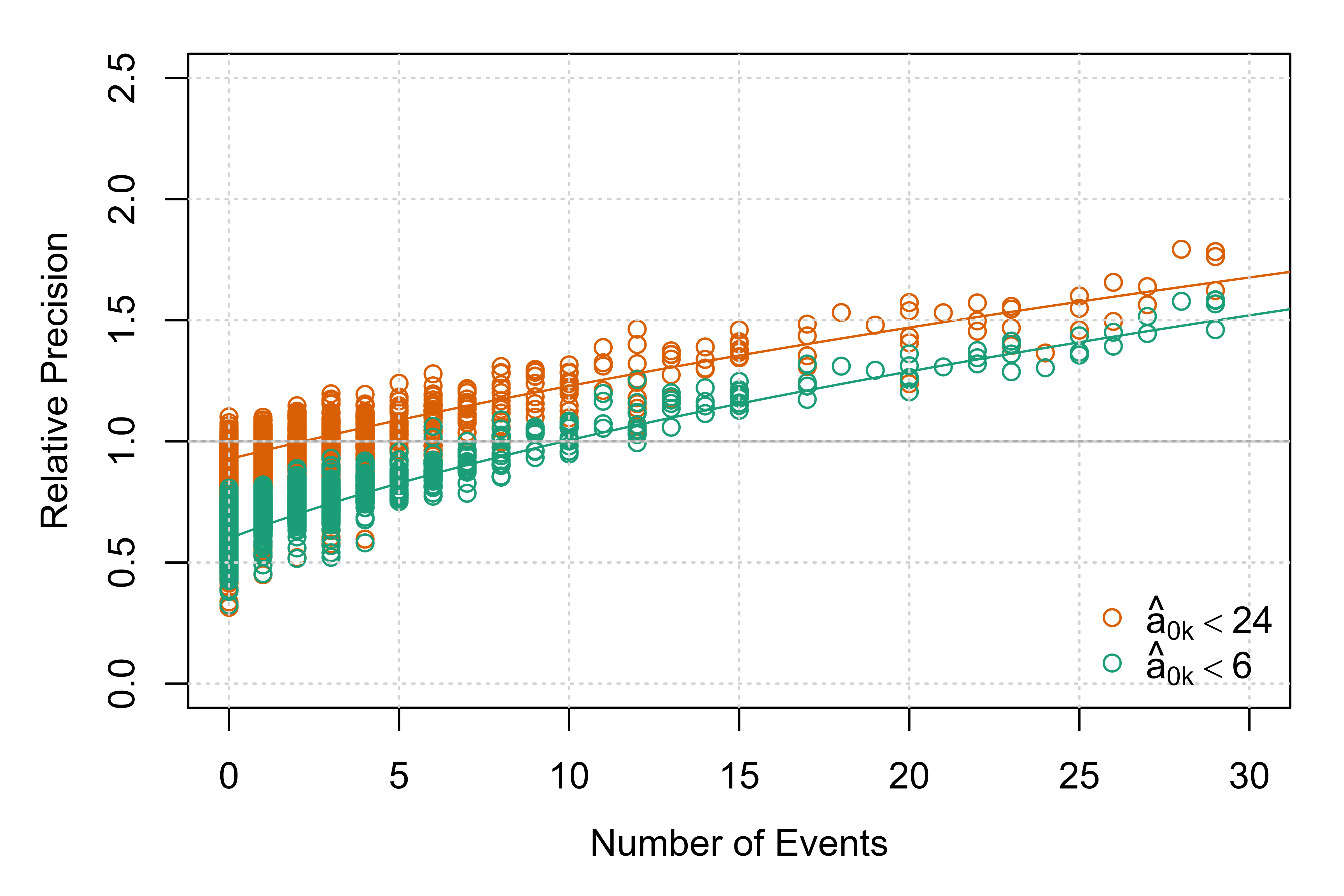}\label{fig:rp_MCAR_bm}}
      \subfigure[MCAR; Black Females]{\includegraphics[width=.45\textwidth]{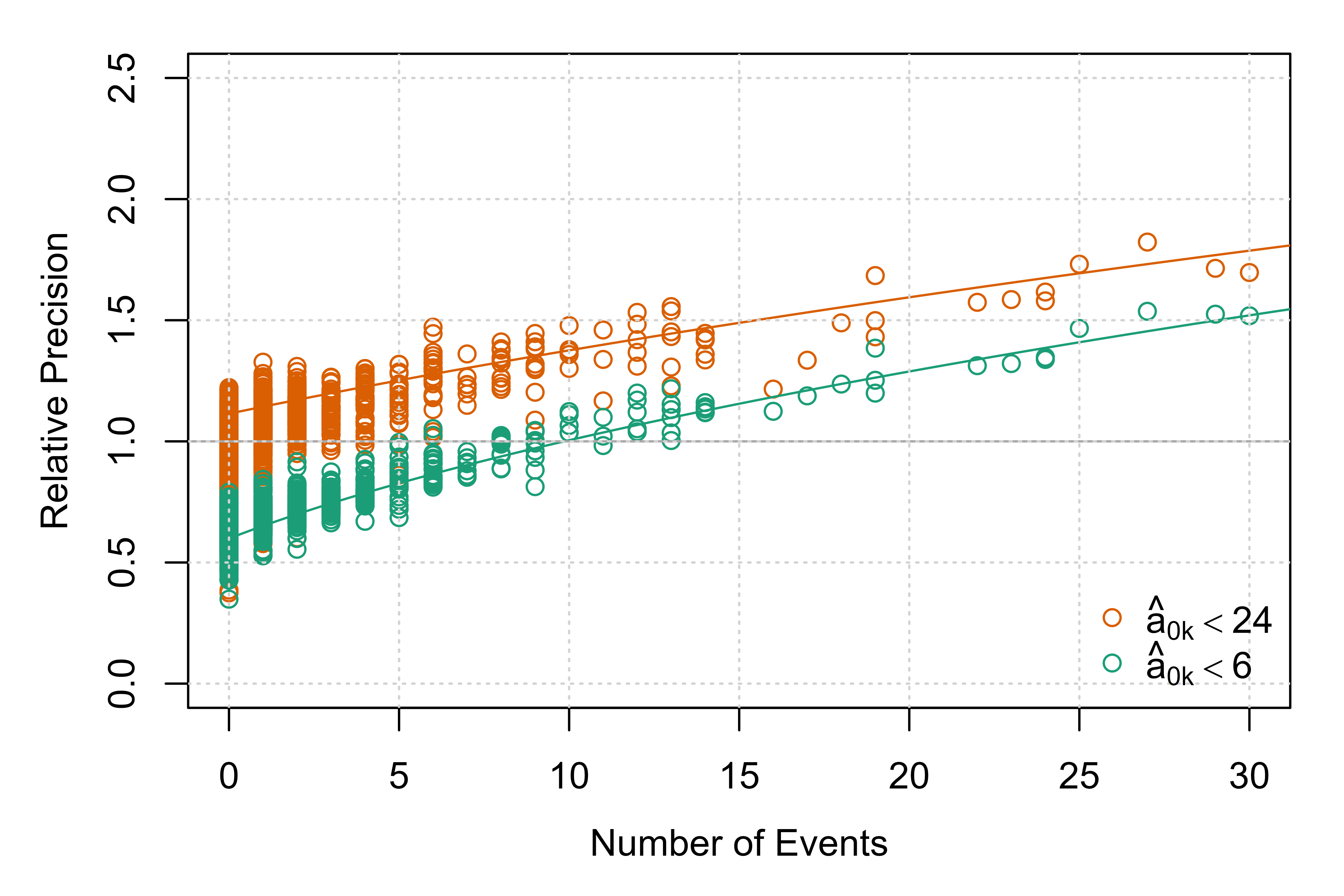}\label{fig:rp_MCAR_bf}}
  \end{center}
  \caption{Relative precision of four groups using two restrictions: $\widehat{a}_{0k} < 24$ and $\widehat{a}_{0k} < 6$}
  \label{fig:rp}
\end{figure}

To further illustrate the extent of oversmoothing---and the resulting artificially narrow credible intervals---in the MCAR estimates, Figure~\ref{fig:rp} compares the relative precision across groups. We consider two thresholds, $\widehat{a}_{0k} < 6$ and $\widehat{a}_{0k} < 24$: the former follows the recommendation of \citet{quick:reliable}, while the latter, four times higher, was chosen to include all groups since the smallest $\widehat{a}_{0k}$ under the unrestricted model was roughly 25 (Black males). Using $A_k = 24$ as an upper benchmark therefore allows all groups to be included in the comparison while still providing a meaningful assessment of the relative precision of the estimates. In Figure~\ref{fig:rp}, we see that when the weaker $\widehat{a}_{0k} < 24$ restriction is imposed, many counties with $y_{ik} = 0$ are deemed reliable based on having a relative precision greater than 1.  In contrast, estimates produced under the more strict $\widehat{a}_{0k} < 6$ restriction fail to achieve the threshold for reliability for most of the counties where $y_{ik} < 10$, as intended. It should also be noted that in Figure~\ref{fig:rp_MCAR_bm} and \ref{fig:rp_MCAR_bf}, the relative precision curves are lower under the $\widehat{a}_{0k} < 24$ restriction for Black males and females compared to White males and females. This is because the estimated informativeness of the standard MCAR model is lower for Black males and females, leading to reduced relative precisions under the same threshold. In contrast, the $\widehat{a}_{0k} < 6$ restriction yields comparable relative precision curves between the two groups, reflecting similar levels of model informativeness.

\begin{figure}[htbp]

  \begin{center}
  \subfigure[CAR; White Male]{\includegraphics[width=.38\textwidth]{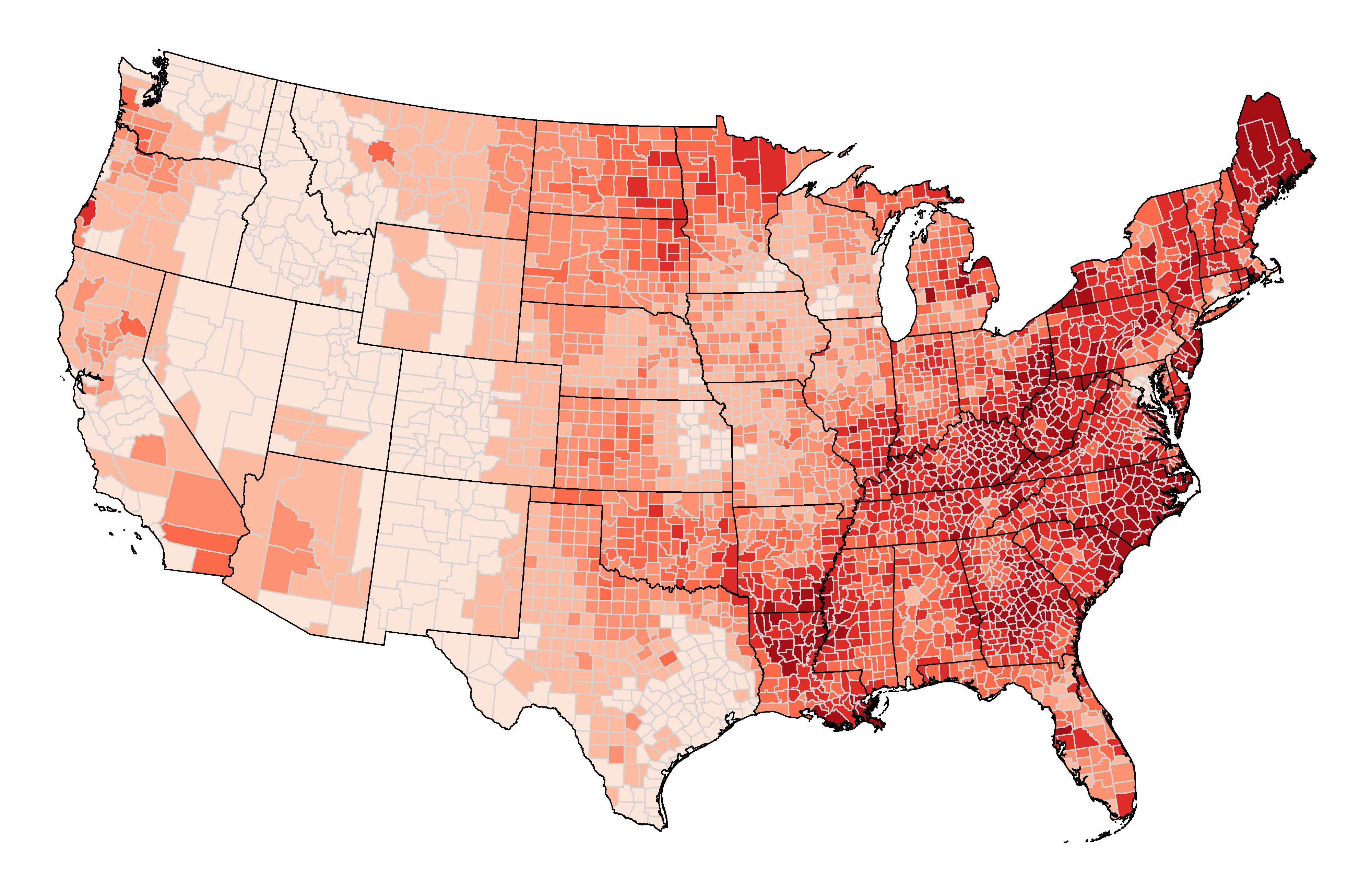}\label{fig:rate1_car_wm}}
  \subfigure[MCAR; White Male]{\includegraphics[width=.38\textwidth]{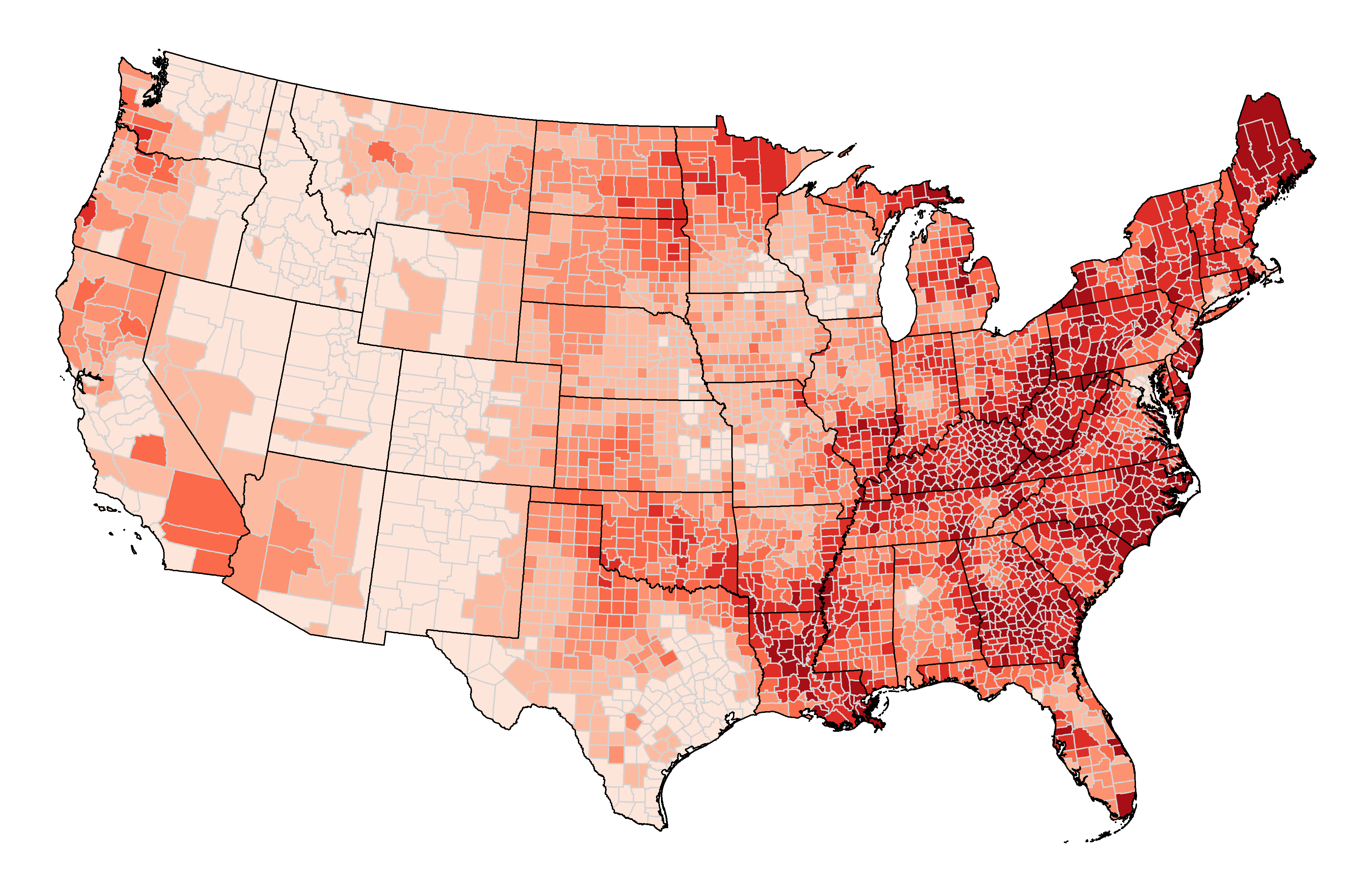}\label{fig:rate1_mcar_wm}}
  \includegraphics[width=.15\textwidth]{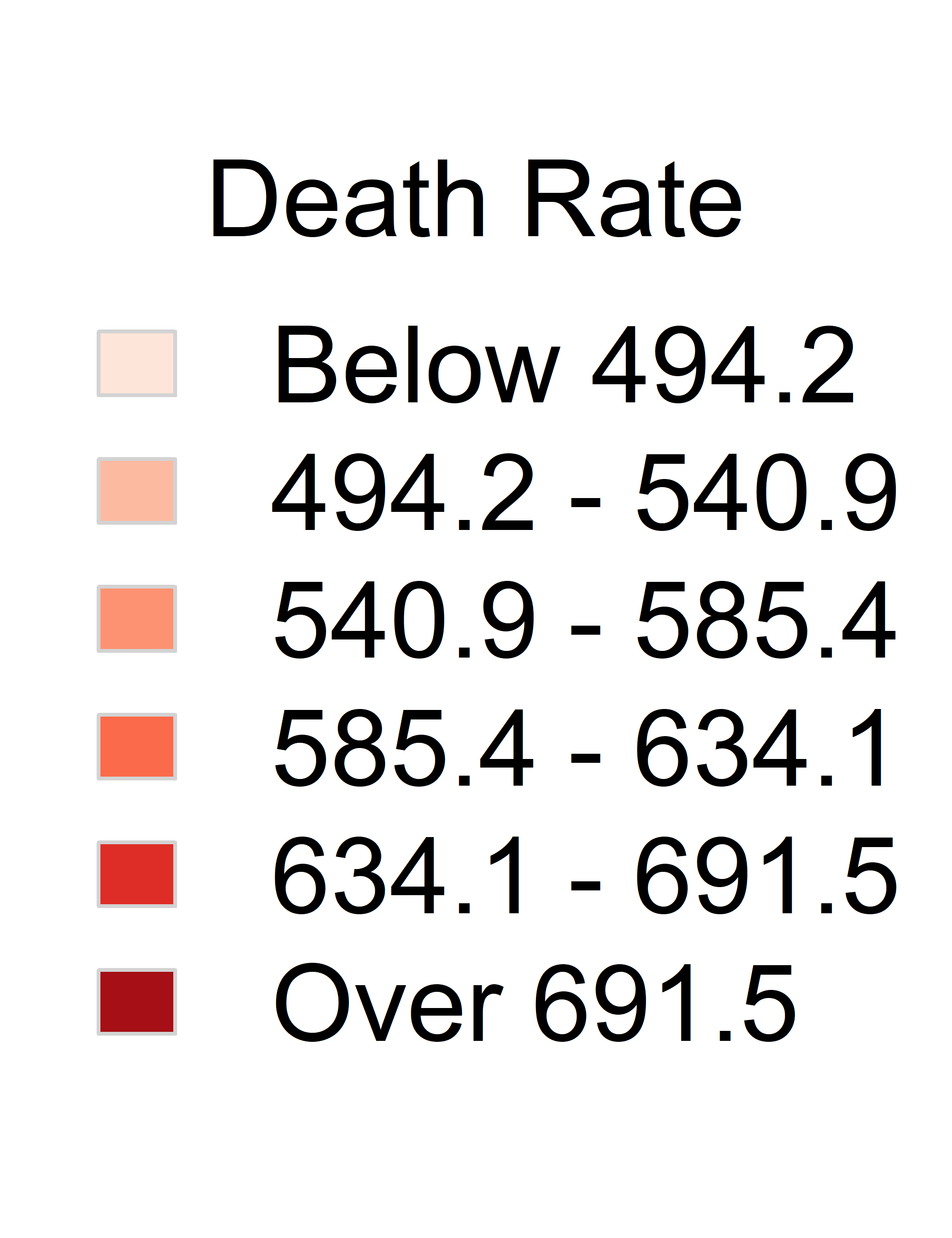}

  \subfigure[CAR; White Female]{\includegraphics[width=.38\textwidth]{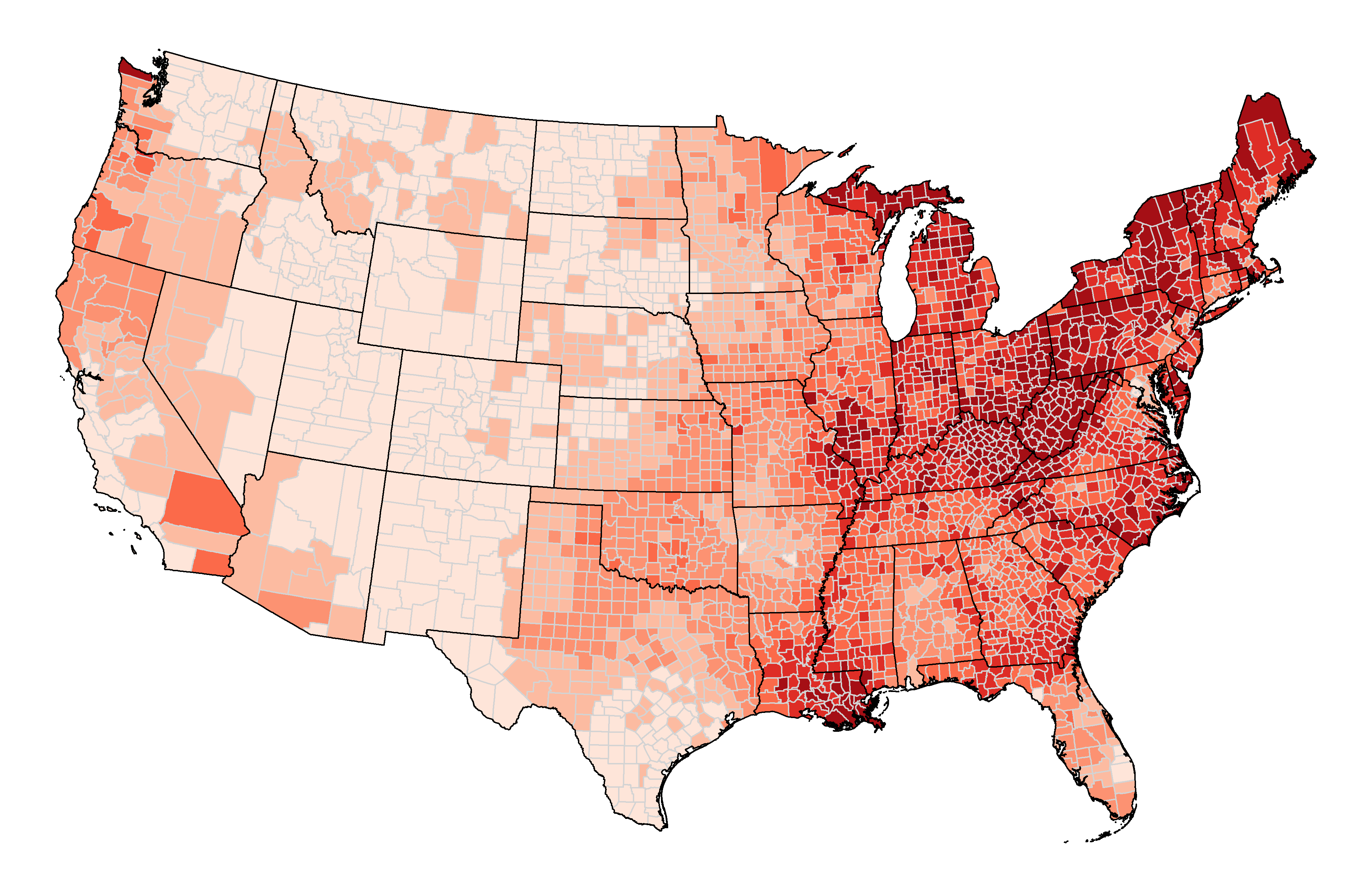}\label{fig:rate1_car_wf}}
  \subfigure[MCAR; White Female]{\includegraphics[width=.38\textwidth]{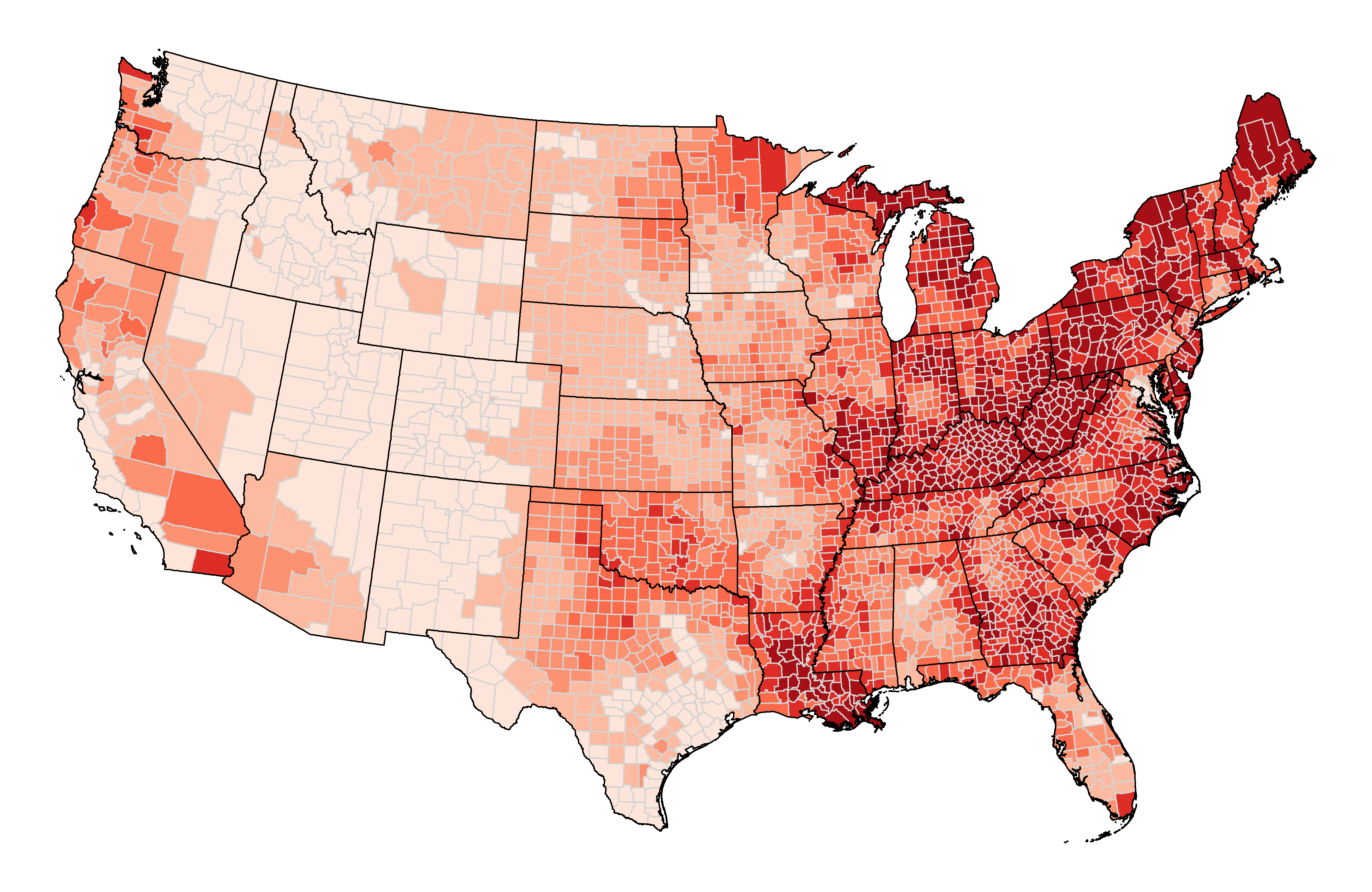}\label{fig:rate1_mcar_wf}}
  \includegraphics[width=.15\textwidth]{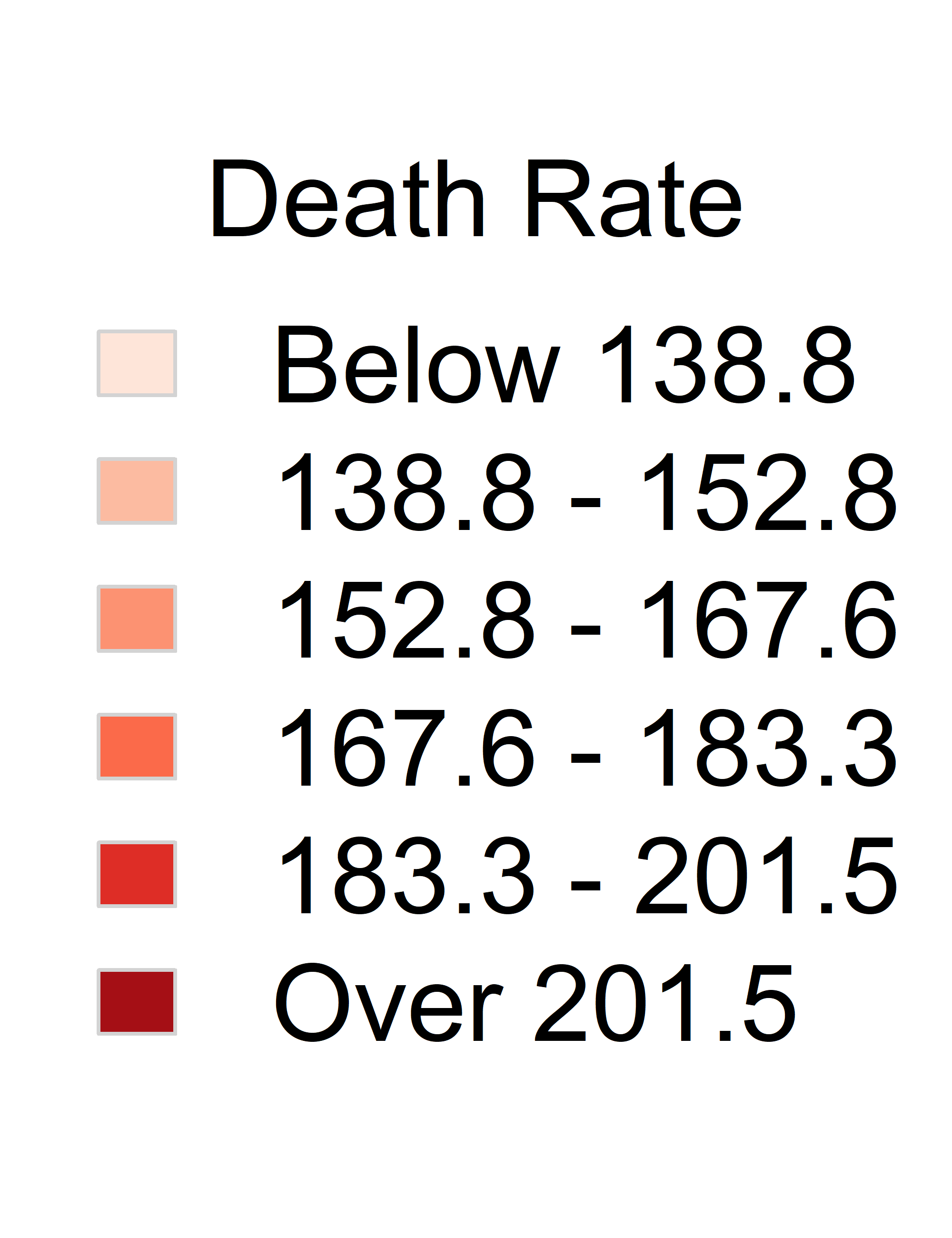}

  \subfigure[CAR; Black Male]{\includegraphics[width=.38\textwidth]{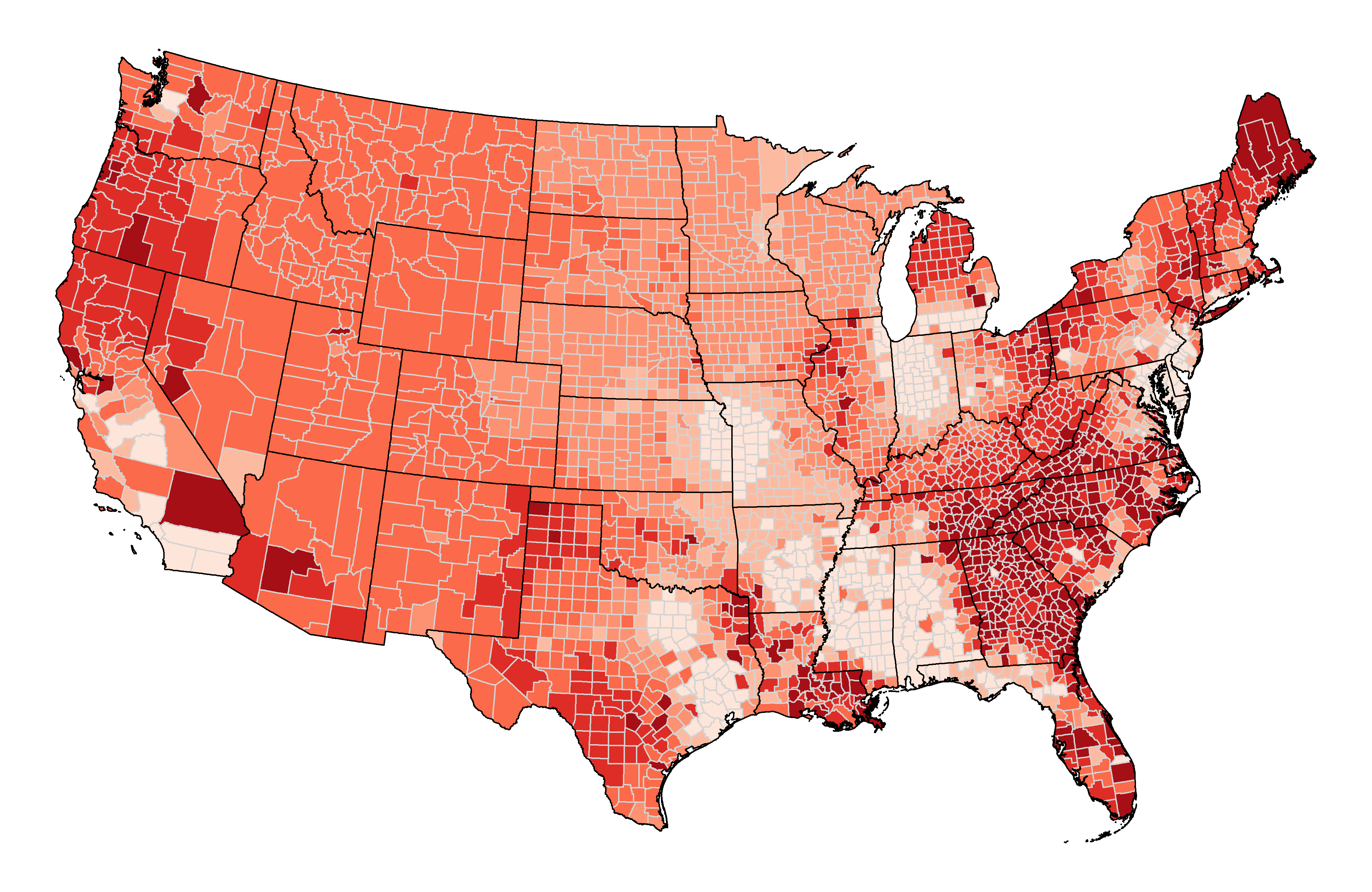}\label{fig:rate1_car_bm}}
  \subfigure[MCAR; Black Male]{\includegraphics[width=.38\textwidth]{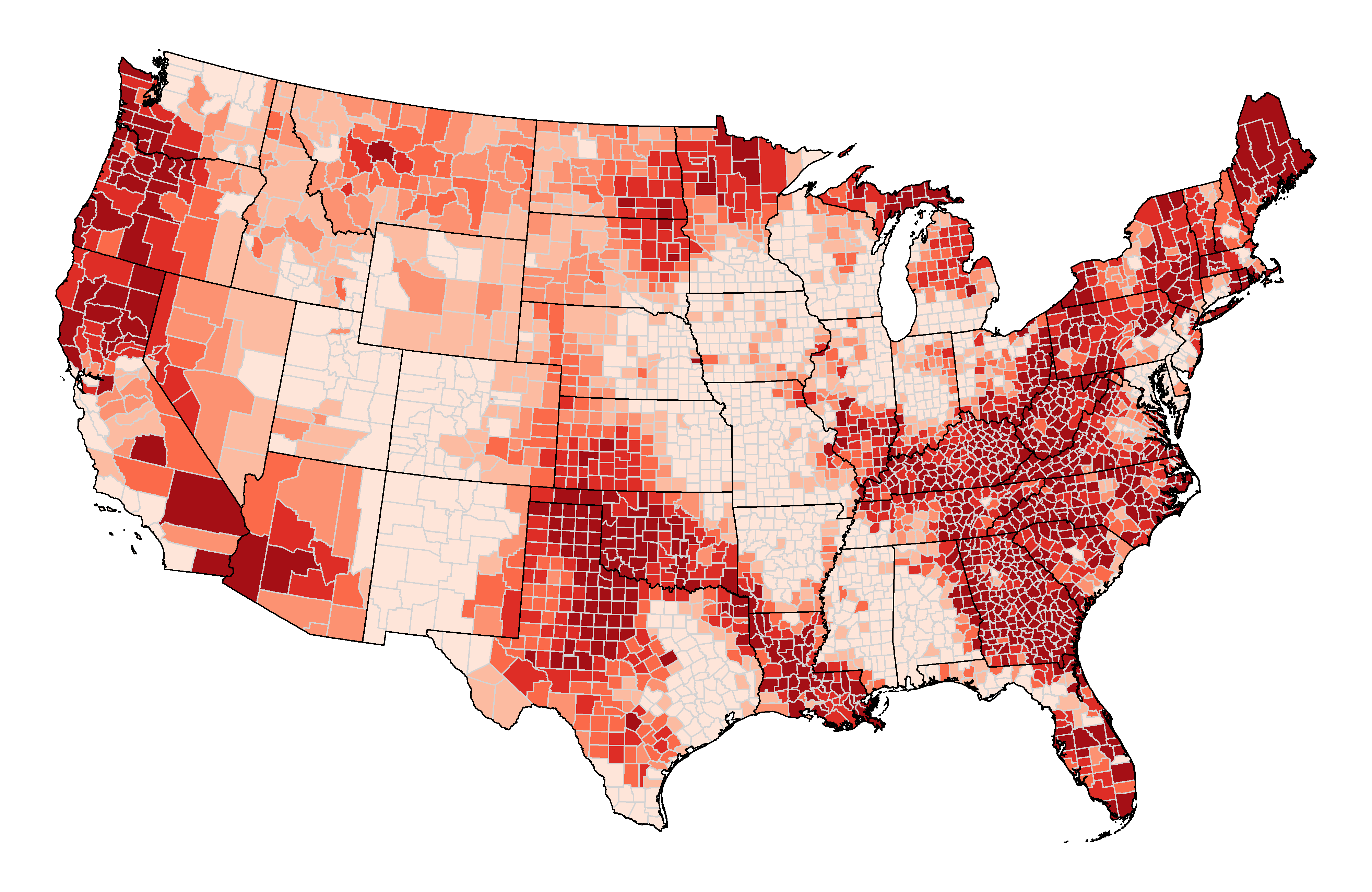}\label{fig:rate1_mcar_bm}}
  \includegraphics[width=.15\textwidth]{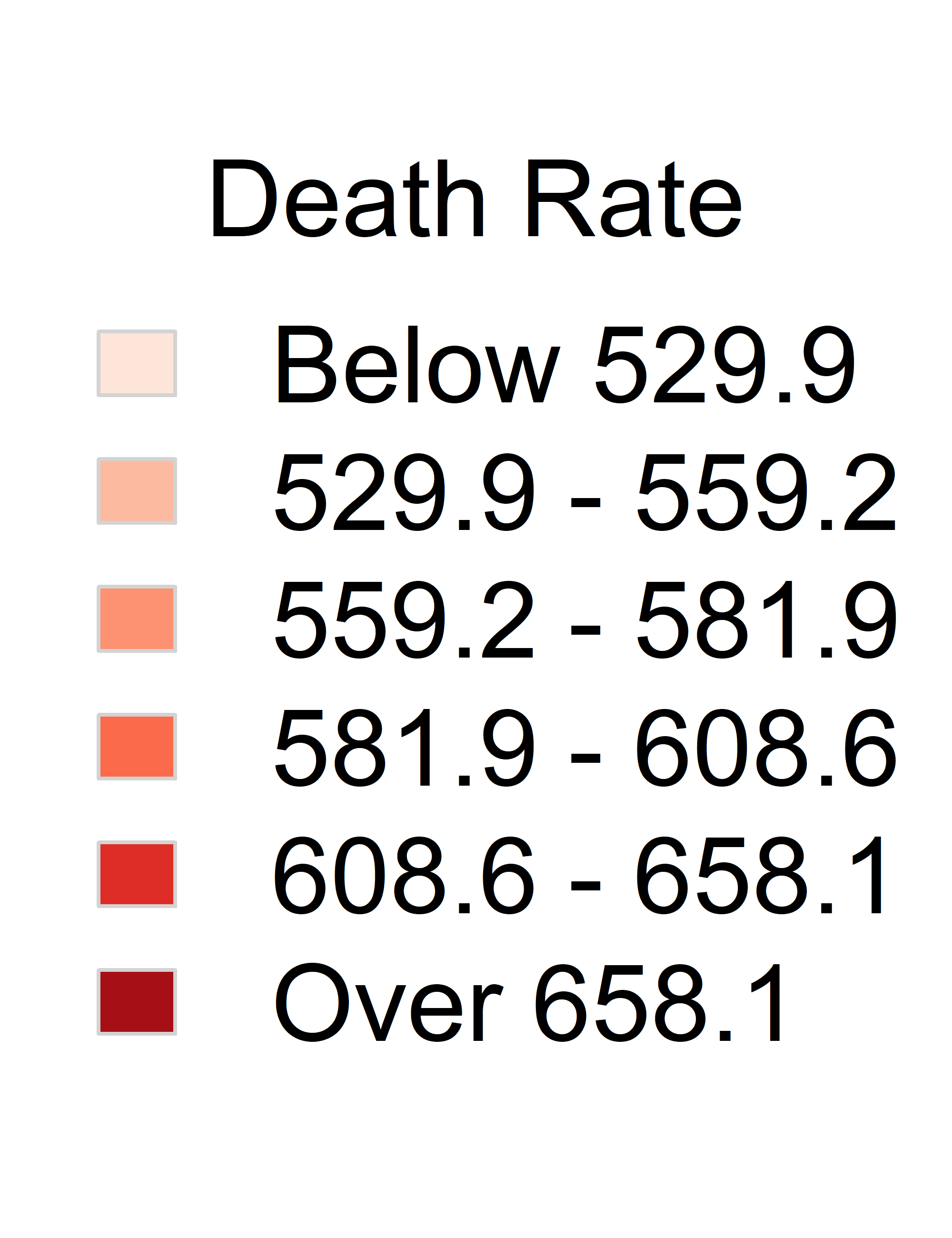}

  \subfigure[CAR; Black Female]{\includegraphics[width=.38\textwidth]{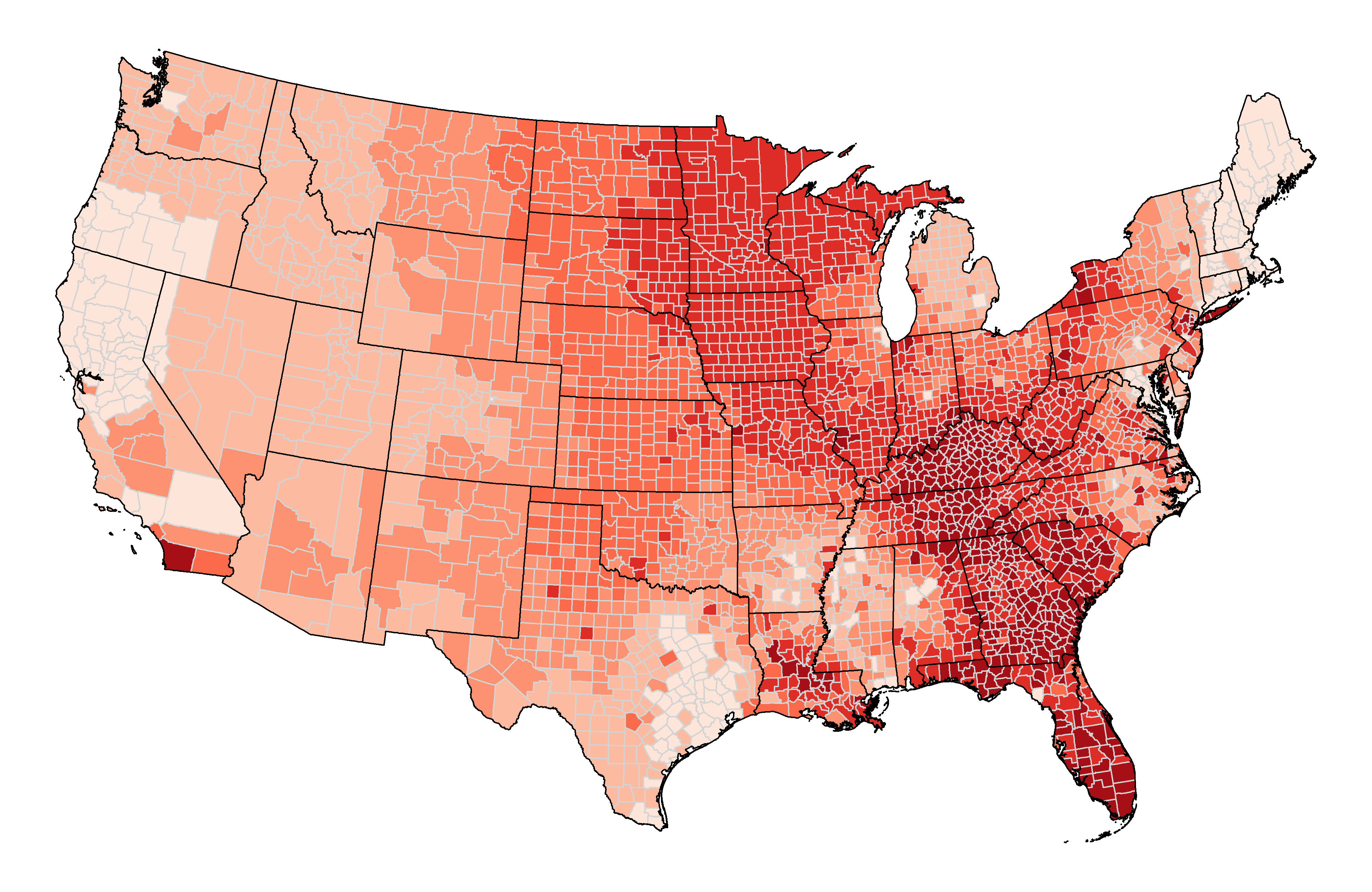}\label{fig:rate1_car_bf}}
  \subfigure[MCAR; Black Female]{\includegraphics[width=.38\textwidth]{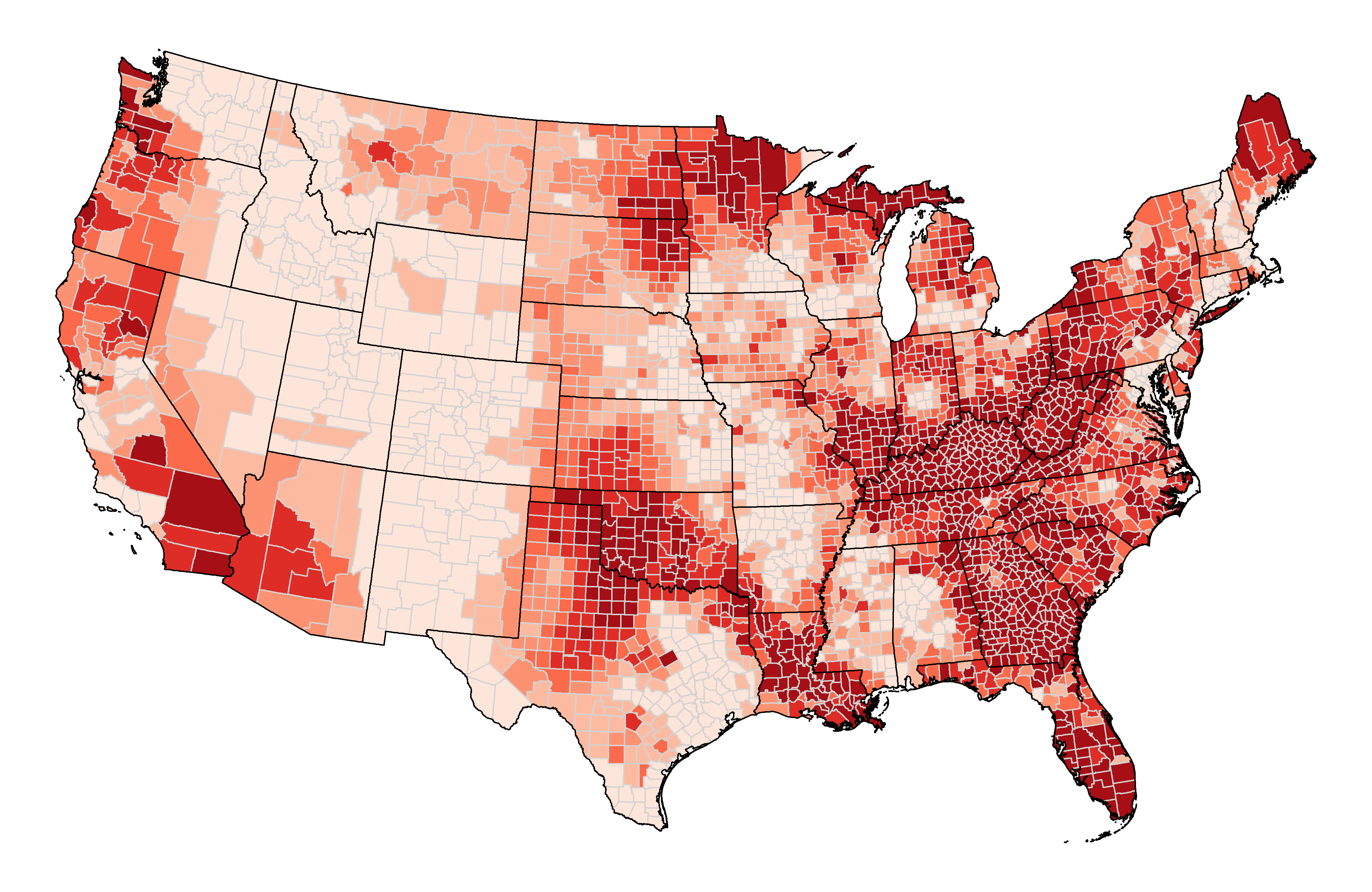}\label{fig:rate1_mcar_bf}}
  \includegraphics[width=.15\textwidth]{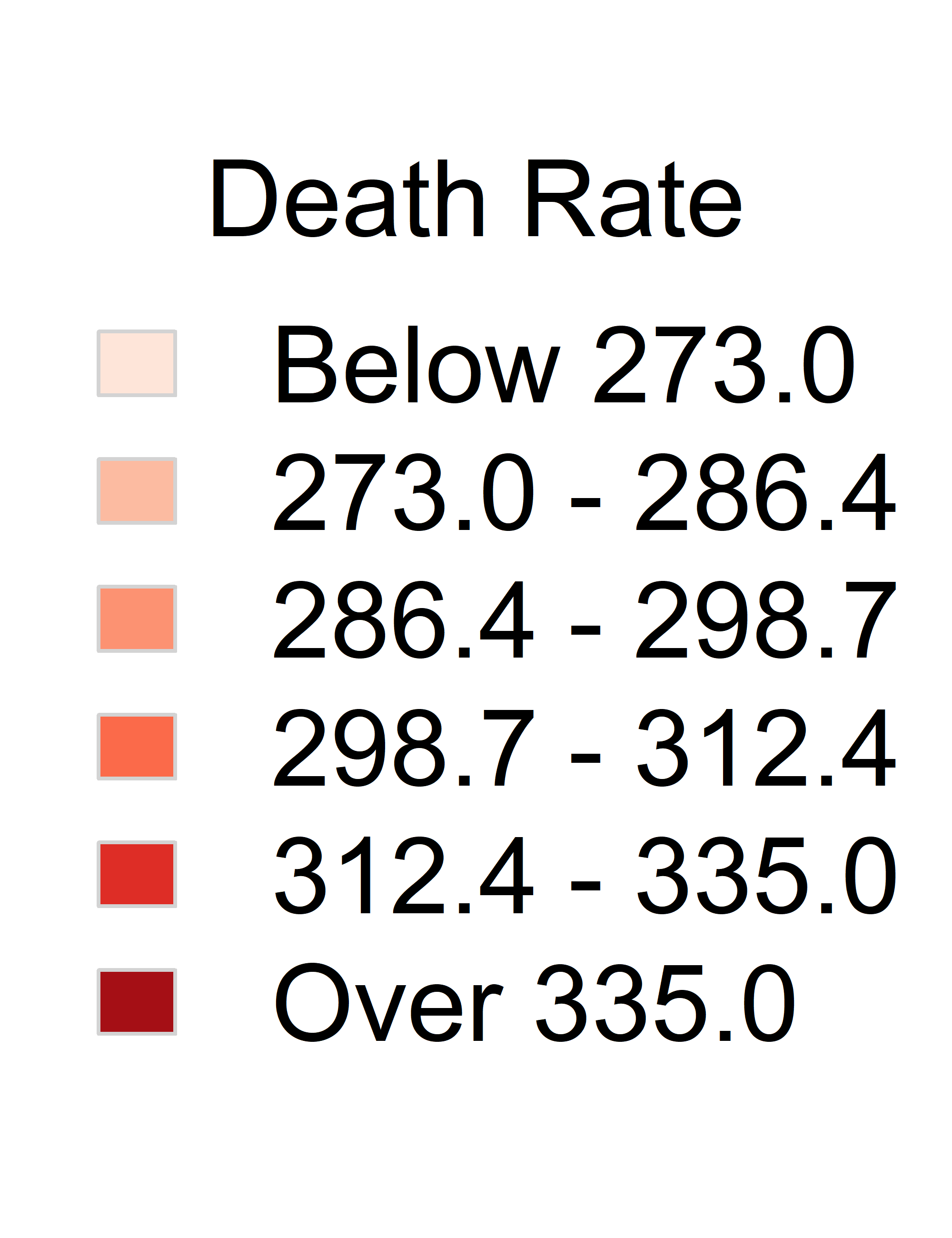}

  \end{center}

  \caption{
    Estimated ischemic heart disease mortality rates per 100,000 using standard CAR and MCAR models. }
  \label{fig:rate1_map}

\end{figure}

Having illustrated the impact of oversmoothing and excessive model informativeness in the context of a \emph{quantitative} measure like relative precision, we now shift to a more \emph{visual} illustration of oversmoothing by examining maps of ischemic heart disease death rates. Figure~\ref{fig:rate1_map} shows the geographical patterns of mortality rates for the four groups using both spatial models. First, we observe that the CAR models (left panel) exhibit overly smooth trends, particularly for Black males and females. For example, the gradient between the West Coast and the Midwest indicates that the model is overpowering the data, as there is insufficient data for these groups to resist the smoothing because the bulk of the Black population resides in coastal areas, urban centers, and the Southeast. In contrast, the MCAR model (right panel) shows a different pattern, as the multivariate structure between the different demographic groups allows them to borrow strength from one another. For example, the MCAR model for Black females shows higher rates along the West Coast (e.g., California and Oregon) and the East Coast (e.g., Maine), as well as lower rates in Utah, Colorado, and New Mexico, compared to the CAR model.  The MCAR model also exhibits sharper transitions between the racially diverse, densely populated urban counties with high rates and their sparsely populated, predominantly White neighboring counties, in contrast to the smoother patterns observed in the CAR model.

\begin{figure}[t]
%Jihyeon used these
%  \subfigure[CAR; Standard]{\includegraphics[width=.4\textwidth]{plots/rate_maps/rate_BM_CAR_standard.png}\label{fig:rate_car_bm}}
%  \subfigure[MCAR; Standard]{\includegraphics[width=.4\textwidth]{plots/rate_maps/rate_BM_MCAR_standard.png}\label{fig:rate_mcar_bm}}
%  \subfigure{\includegraphics[width=.15\textwidth]{plots/rate_maps/rate_leg_BM.png}\label{fig:rate_leg_bm}}

%  \setcounter{subfigure}{2}

  % \subfigure[CAR; $\widehat{a}_{0k} < 30$]{\includegraphics[width=.4\textwidth]{plots/rate_maps/rate_BM_CAR_5.png}\label{fig:rate_car6_bm2}}
  % \subfigure[MCAR; $\widehat{a}_{0k} < 30$]{\includegraphics[width=.4\textwidth]{plots/rate_maps/rate_BM_MCAR_5.png}\label{fig:rate_mcar6_bm2}}
  % \subfigure{\includegraphics[width=.15\textwidth]{plots/rate_maps/rate_leg_BM.png}\label{fig:rate_leg_bm2}}

  % \setcounter{subfigure}{4}

  \subfigure[CAR; $\widehat{a}_{0k} < 6$]{\includegraphics[width=.4\textwidth]{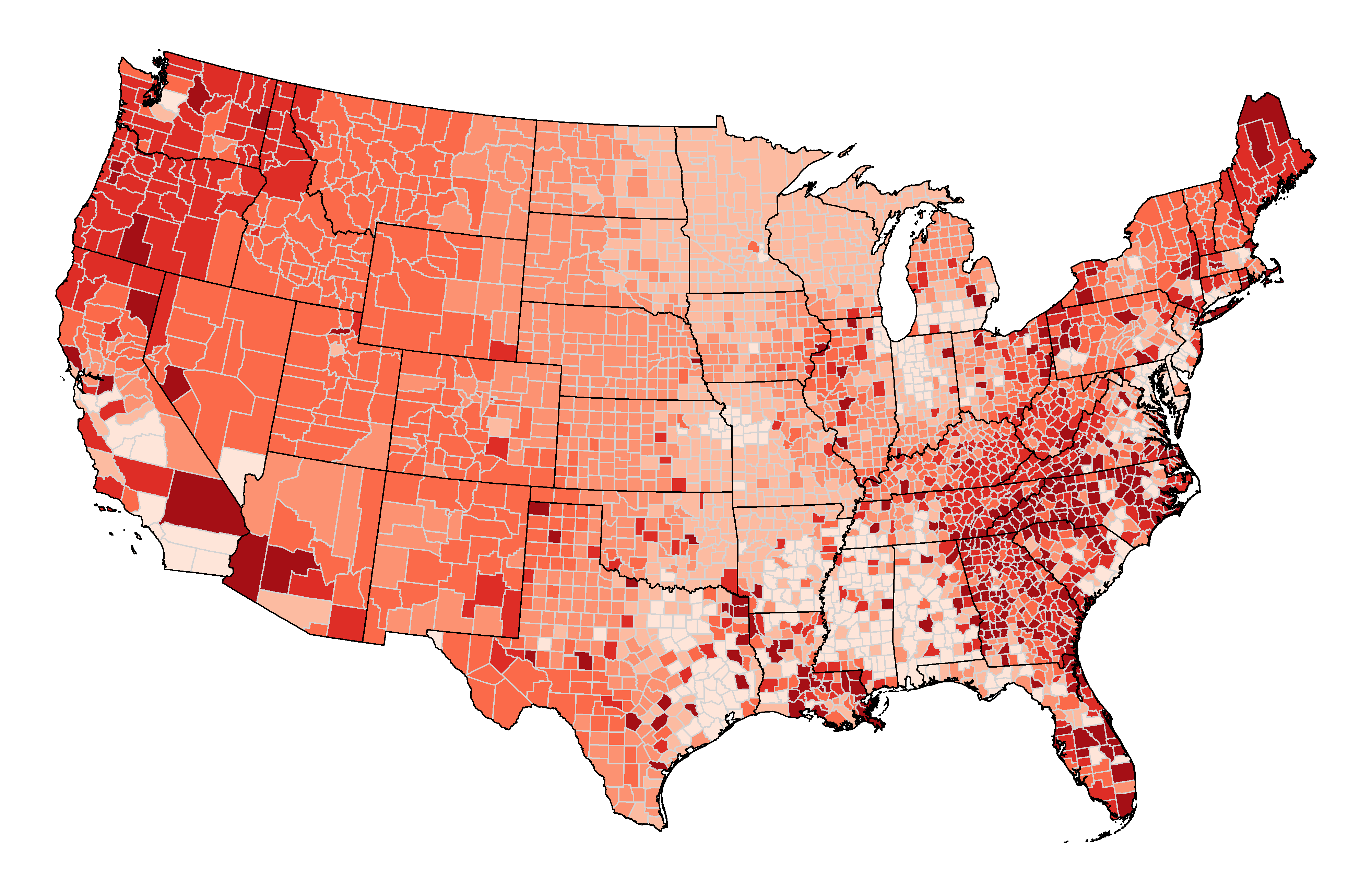}\label{fig:rate_car6_bm2}}
  \subfigure[MCAR; $\widehat{a}_{0k} < 6$]{\includegraphics[width=.4\textwidth]{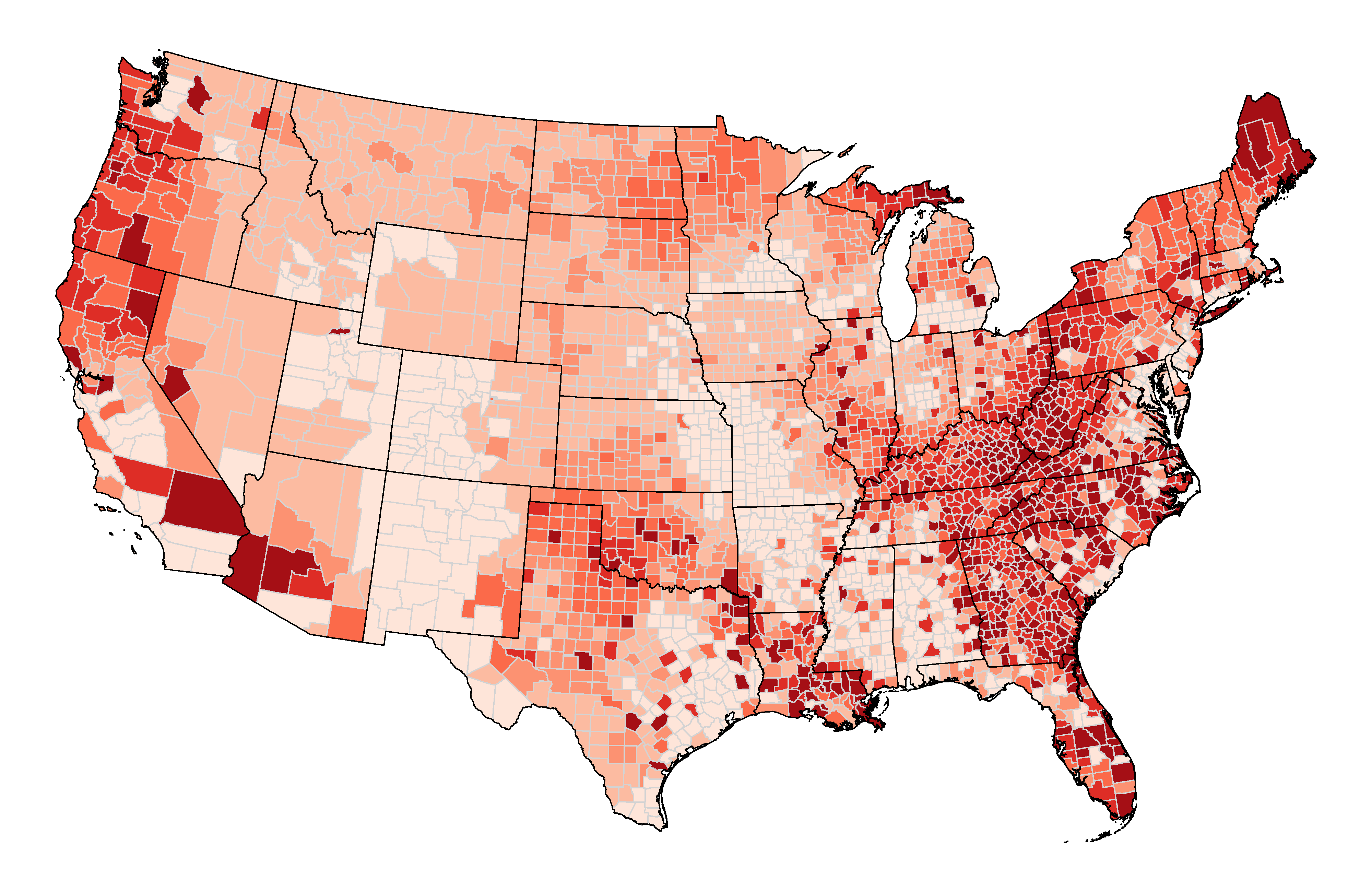}\label{fig:rate_mcar6_bm2}}
  \includegraphics[width=.15\textwidth]{plots/rate_maps/rate_leg_BM.png}

  \caption{
    Estimated ischemic heart disease mortality rates per 100,000 for Black males using CAR and MCAR models under the informativeness restriction of $\hat{a}_{0k} < 6$.
    }
  \label{fig:rate_map_bm2}

\end{figure}

Figure~\ref{fig:rate_map_bm2} compares the estimates from the two restricted ($\widehat{a}_{0k} < 6$) models for Black males (see Appendix~C for other groups and thresholds). Here, we see that the differences in the estimates from the two restricted models are less stark than they were in the unrestricted models in Figures~\ref{fig:rate1_car_bm} and~\ref{fig:rate1_mcar_bm}, but the differences nevertheless highlight the benefits of the MCAR framework.  In particular, whereas the CAR model's estimates in Figure~\ref{fig:rate_car6_bm2} exhibit the same smooth gradient between the West Coast and the Midwest as observed in Figure~\ref{fig:rate1_car_bm} due to the relative lack of data from Black males throughout many of these counties, the restricted MCAR still allows the estimates for Black men to learn from the estimates for White men---albeit to a lesser extent than under the standard MCAR---thereby producing geographic patterns that are more reminiscent of those observed in Figure~\ref{fig:rate1_car_wm}.
%We observe that, compared to the restricted CAR model, the restricted MCAR model tends to produce lower rates in {\color{blue}much of the Mountain West region}, %Seattle, Oregon, and the Mountain areas,
%while also showing clusters of counties with higher rates than surroundings, such as in {\color{blue}states like} Minnesota.
%In contrast, the restricted CAR model yielded a more gradual transition from the West Coast to the Midwest, with a smoother pattern overall.
%This demonstrates that with the MCAR model, we obtain different geographical patterns due to the dependency structure being considered, even with the same amount of information.

\begin{figure}[htbp]

  \begin{center}
  \subfigure[Standard; White Male]{\includegraphics[width=.38\textwidth]{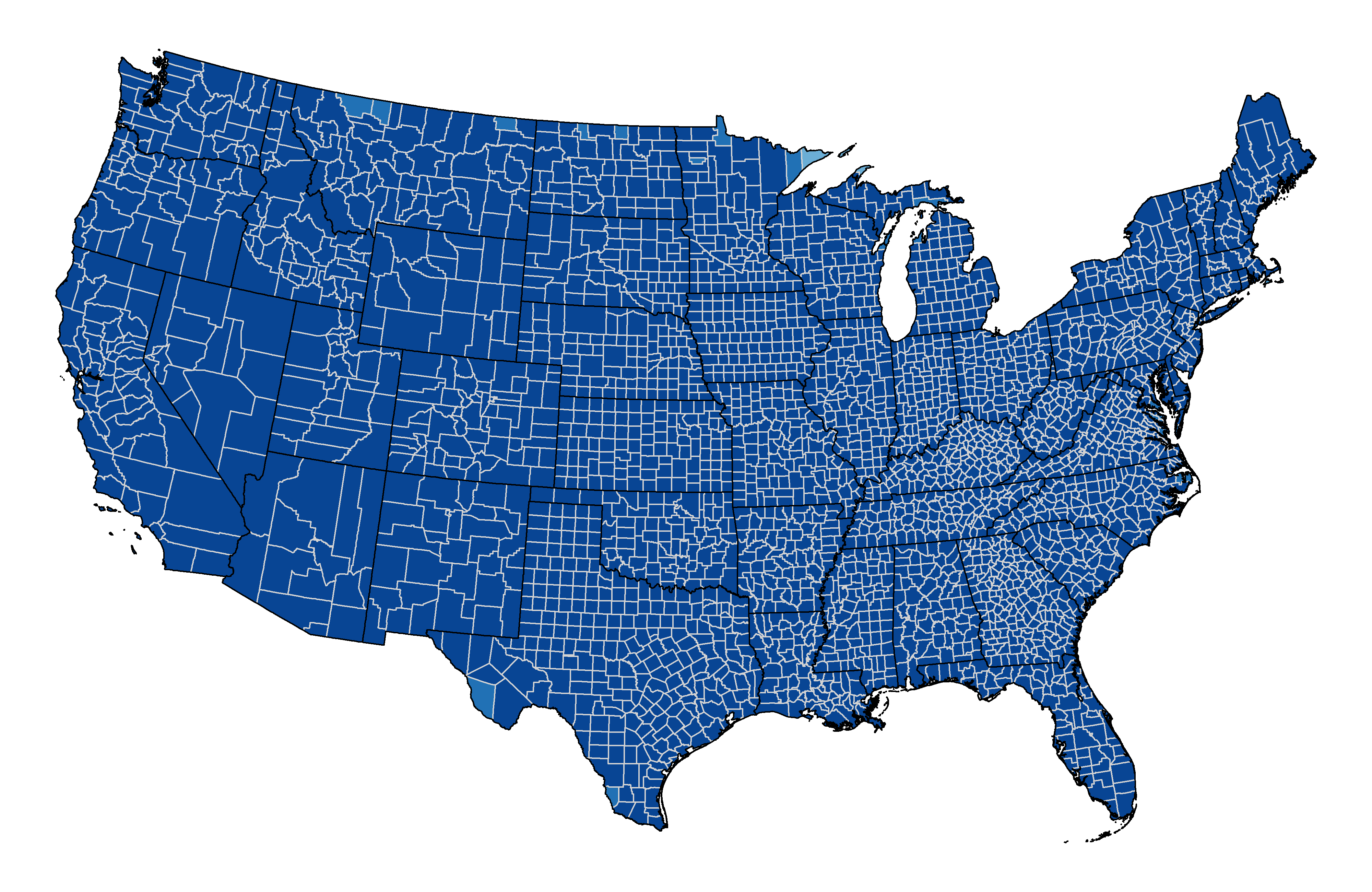}\label{fig:rel_wm}}
  \subfigure[$\widehat{a}_{0k} < 6$; White Male]{\includegraphics[width=.38\textwidth]{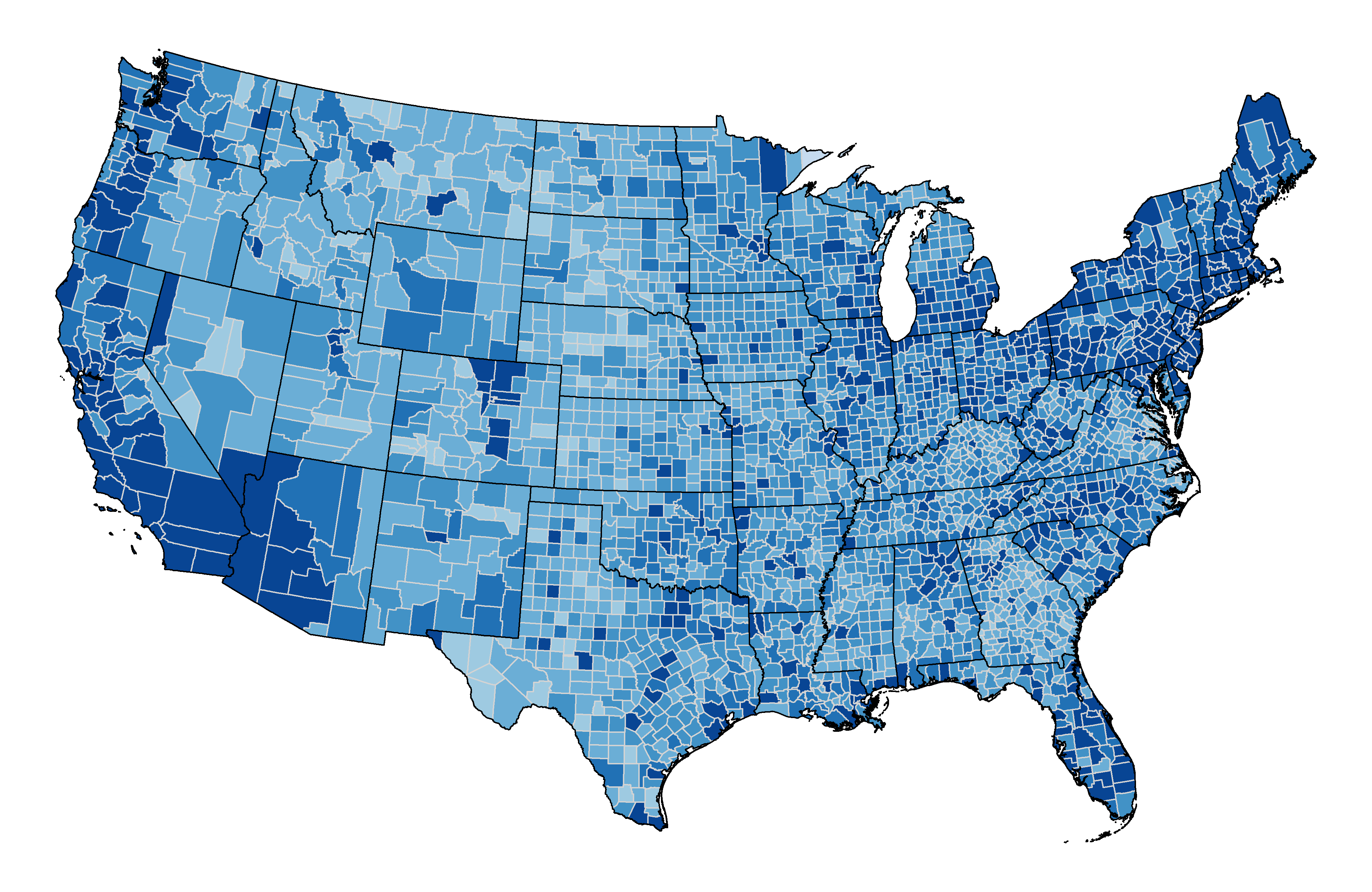}\label{fig:rel_res_wm}}
  \includegraphics[width=.15\textwidth]{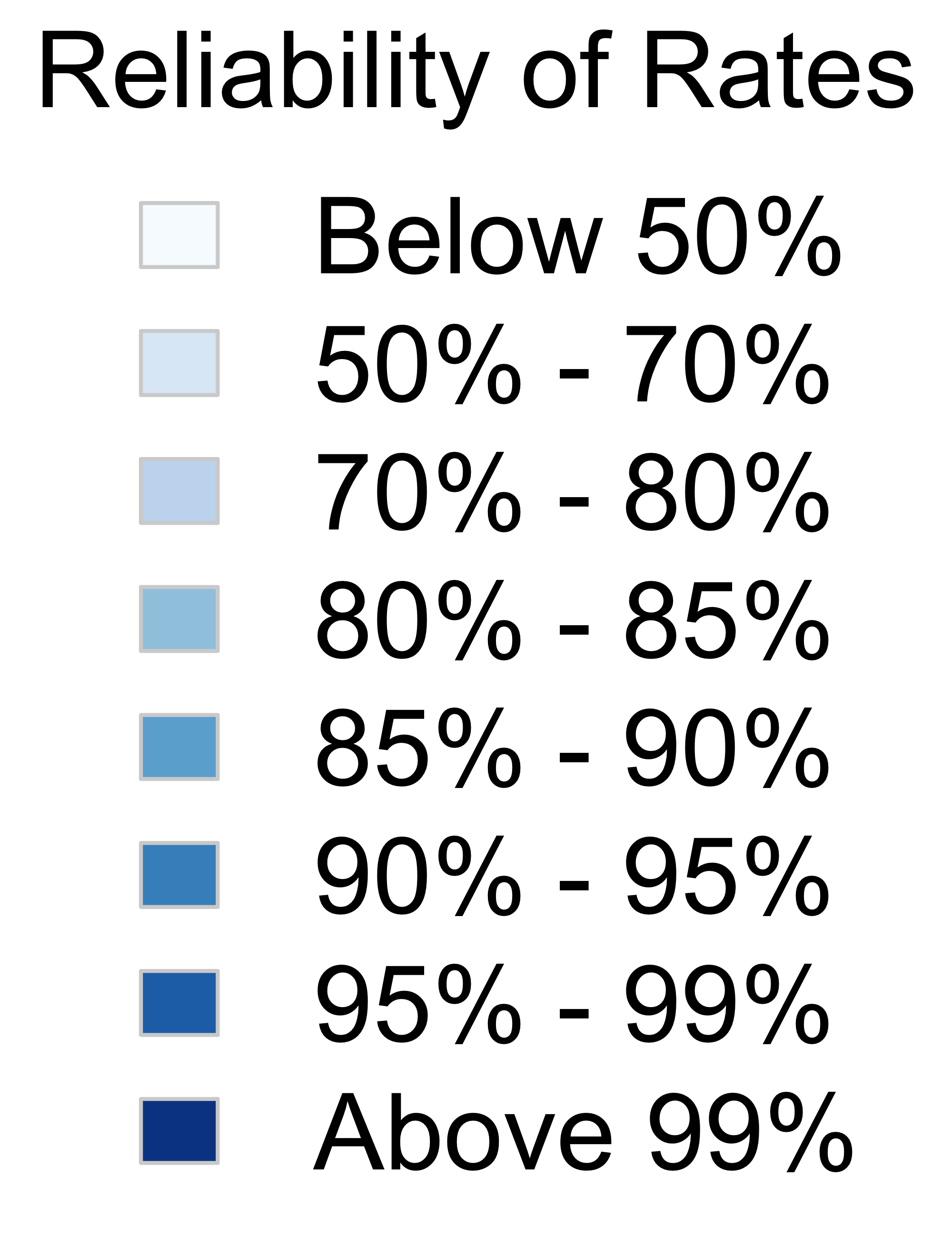}

  \subfigure[Standard; White Female]{\includegraphics[width=.38\textwidth]{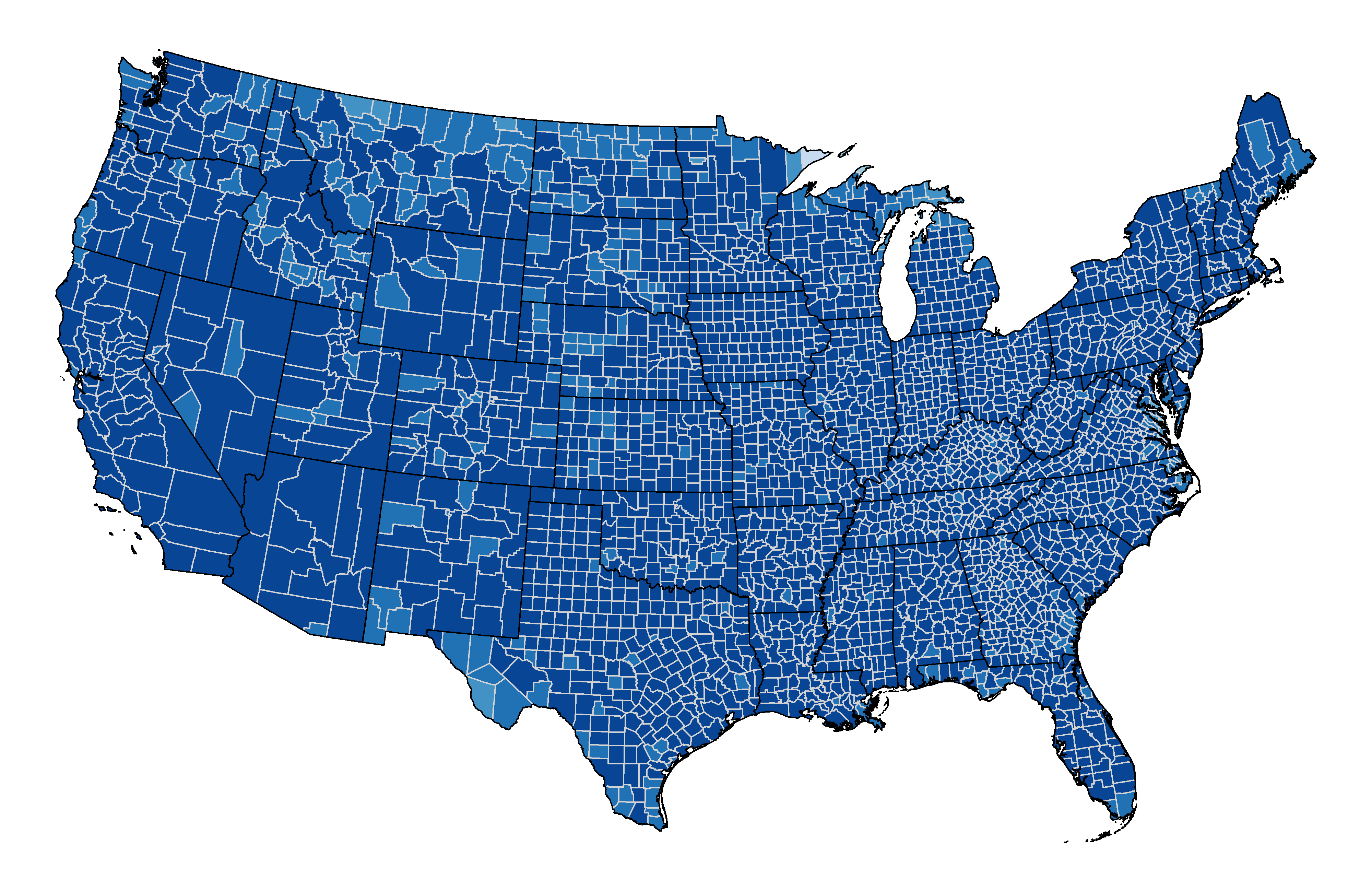}\label{fig:rel_wf}}
  \subfigure[$\widehat{a}_{0k} < 6$; White Female]{\includegraphics[width=.38\textwidth]{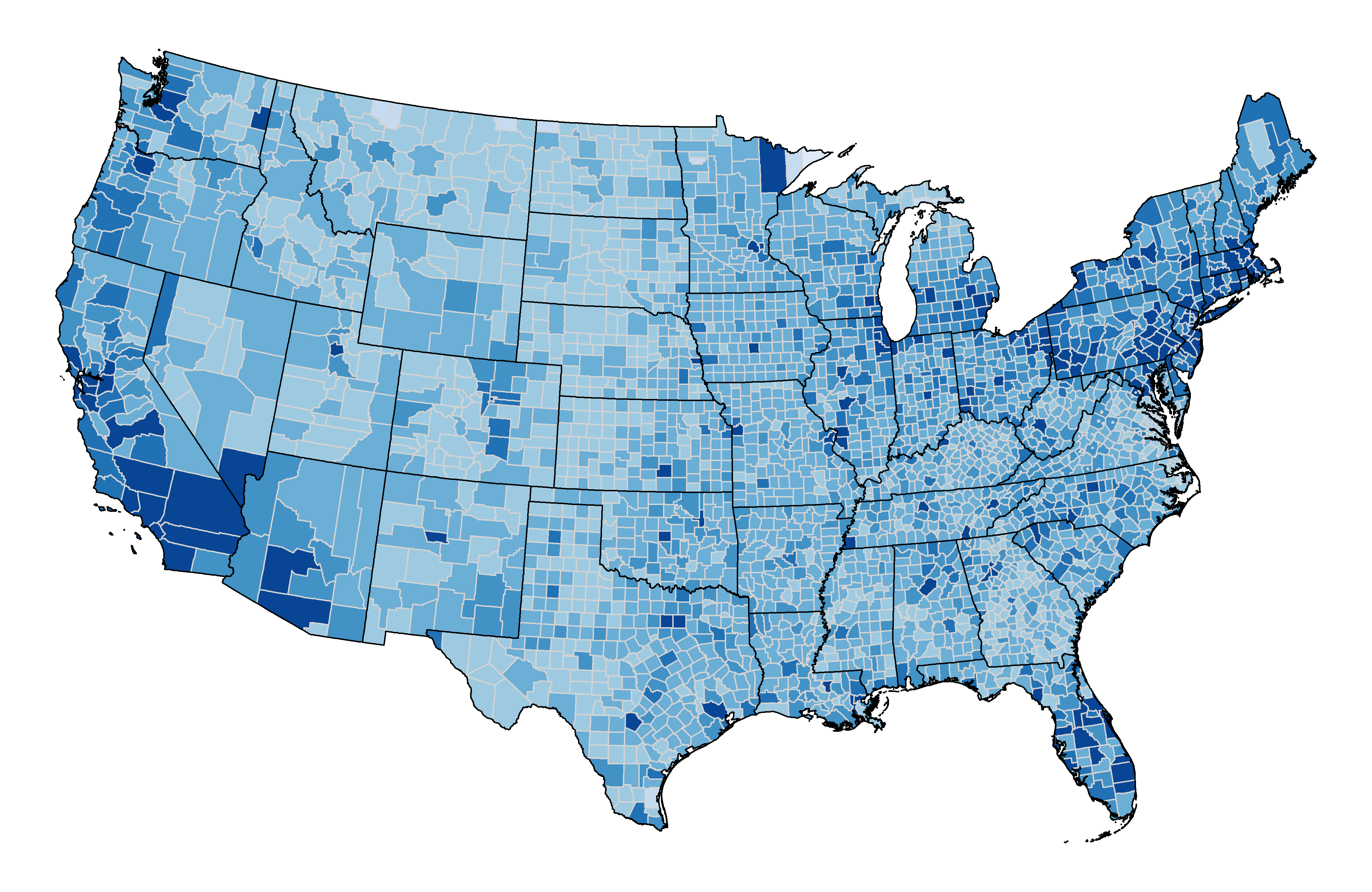}\label{fig:rel_res_wf}}
  \includegraphics[width=.15\textwidth]{plots/rate_maps/rel_leg.png}

  \subfigure[Standard; Black Male]{\includegraphics[width=.38\textwidth]{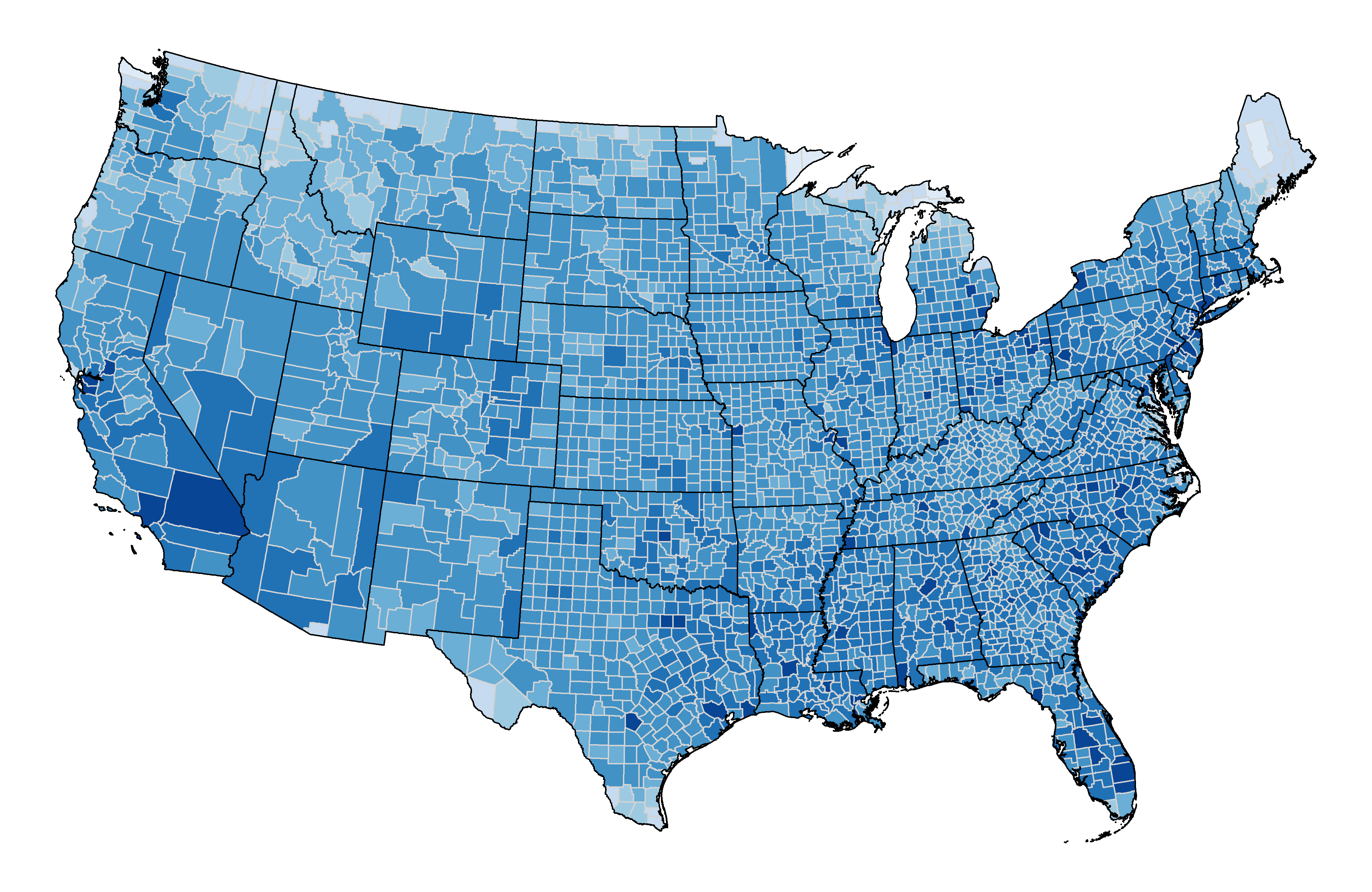}\label{fig:rel_bm}}
  \subfigure[$\widehat{a}_{0k} < 6$; Black Male]{\includegraphics[width=.38\textwidth]{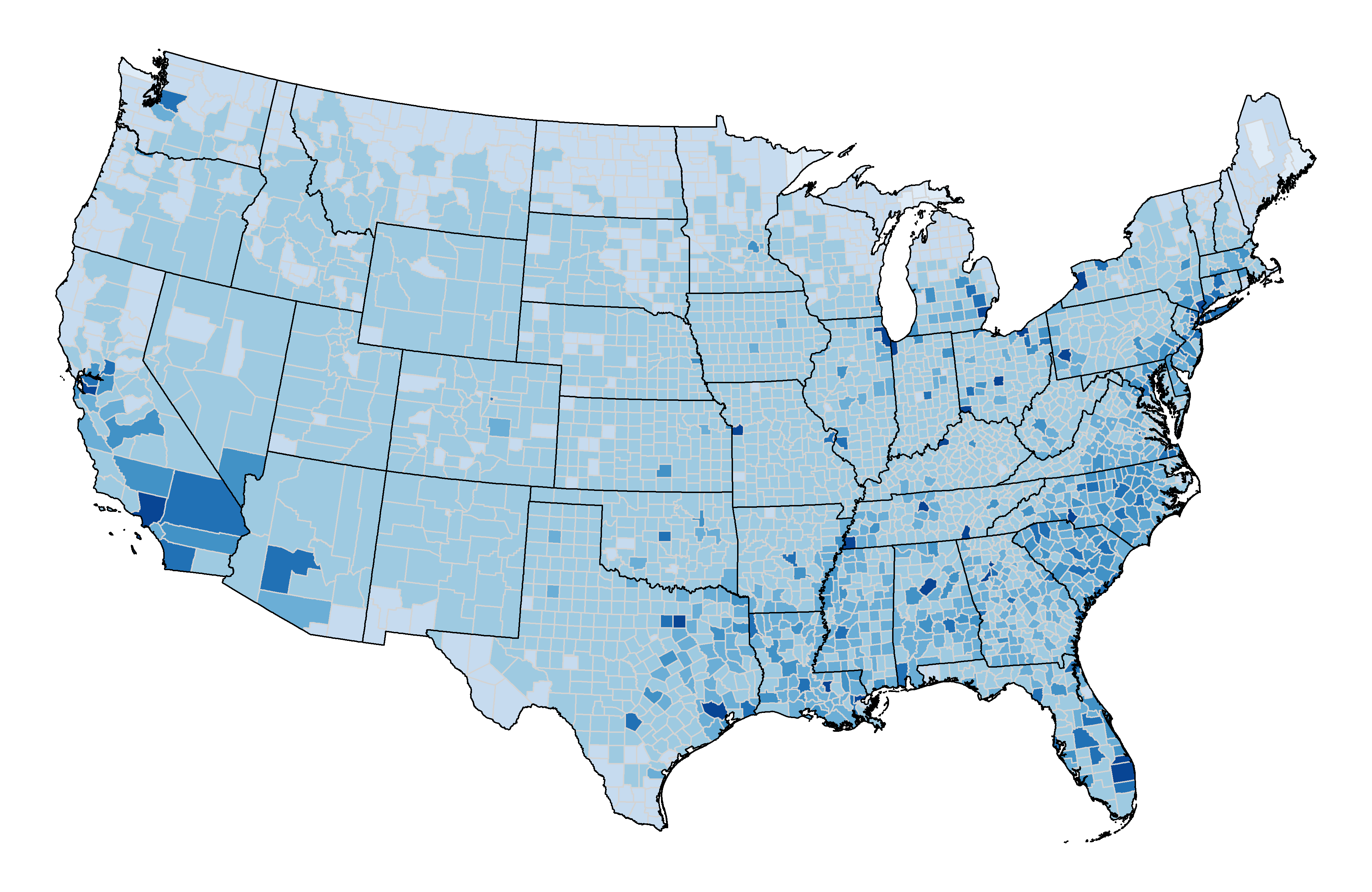}\label{fig:rel_res_bm}}
  \includegraphics[width=.15\textwidth]{plots/rate_maps/rel_leg.png}

  \subfigure[Standard; Black Female]{\includegraphics[width=.38\textwidth]{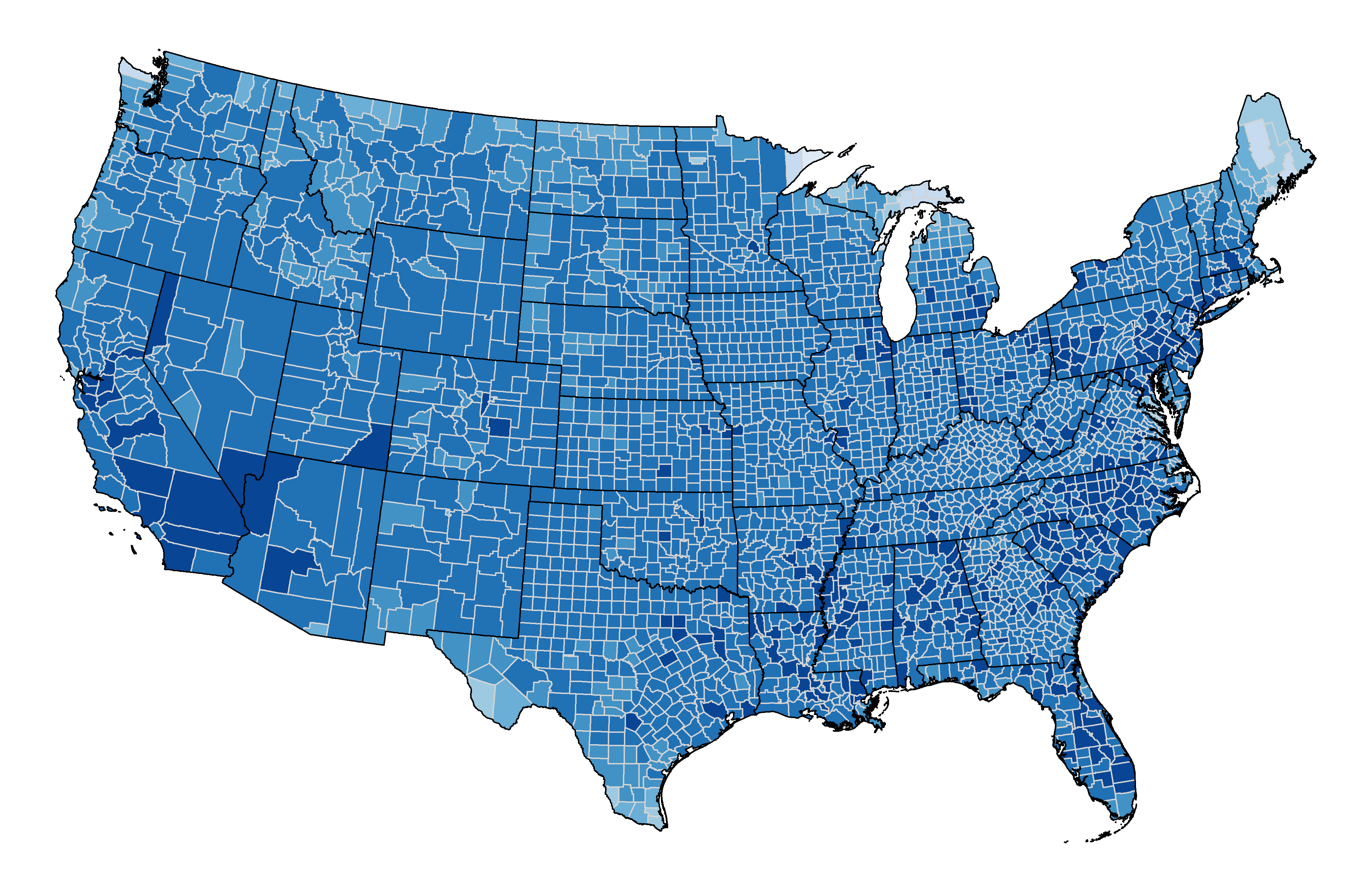}\label{fig:rel_bf}}
  \subfigure[$\widehat{a}_{0k} < 6$; Black Female]{\includegraphics[width=.38\textwidth]{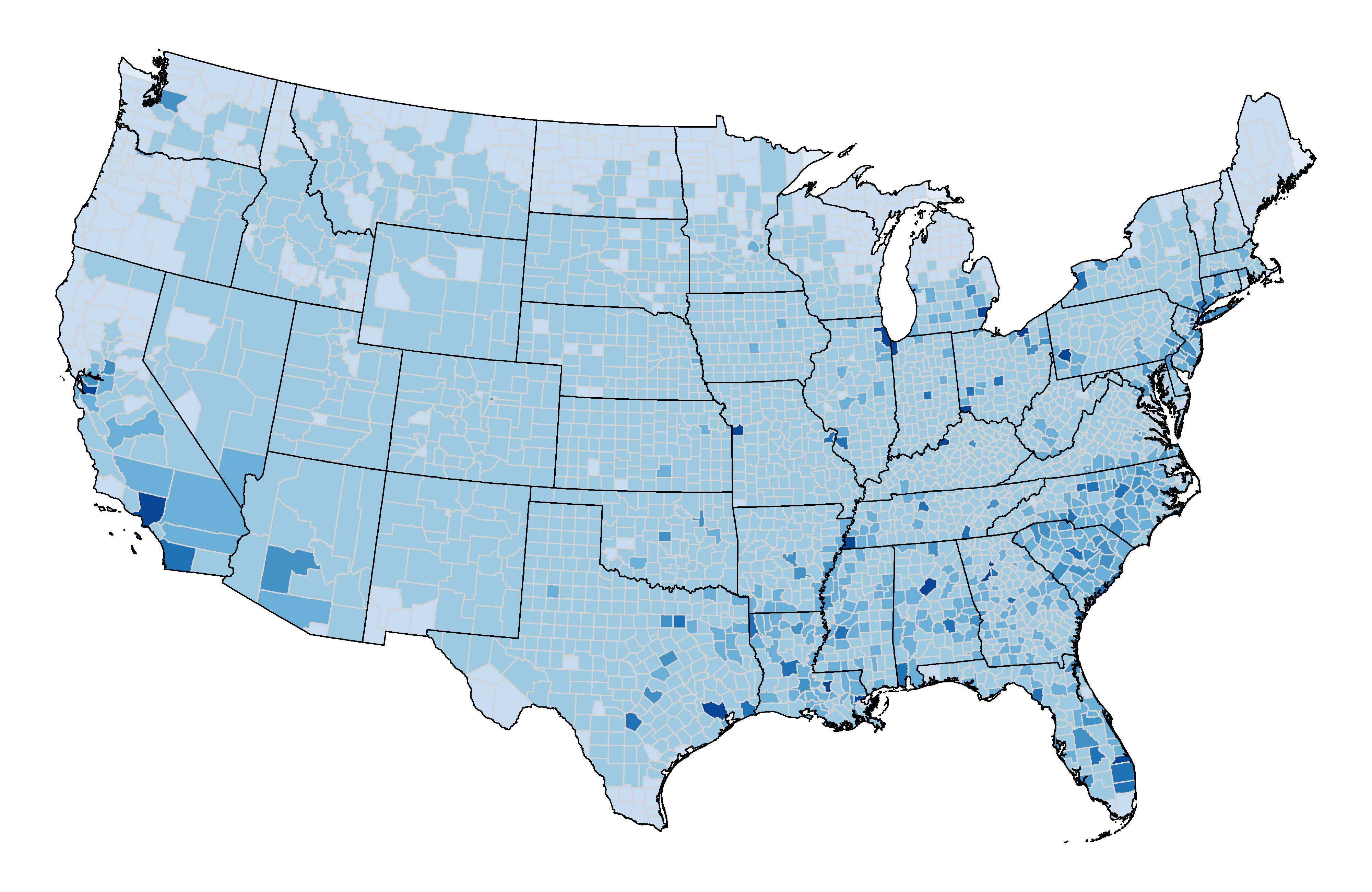}\label{fig:rel_res_bf}}
  \includegraphics[width=.15\textwidth]{plots/rate_maps/rel_leg.png}
  \end{center}

  \caption{
    Reliability of rates using the standard and restricted MCAR models.
    }
  \label{fig:rel_map}

\end{figure}

While the differences between the CAR and MCAR model estimates in Figure~\ref{fig:rate_map_bm2} in sparsely populated areas are noteworthy, it is important to keep in mind that reducing the informativeness of the model affects both the point estimates we display in maps \textit{and} the precision of those estimates.  Thus, it is also important to assess the reliability of rate estimates to help focus attention on the areas where we have the most confidence.  To that end,
%ensure that artificially reliable estimates do not arise due to high model informativeness when little to no data is available.
Figure~\ref{fig:rel_map} illustrates the reliability of the rate estimates under the standard and restricted ($\widehat{a}_{0k}<6$) models for each subgroup, measured as the largest $\left(1-\alpha\right)\times 100\%$ CI whose width is less than the posterior median, per \citet{quick:reliable}.
Here, the impact of the standard MCAR model's excessive informativeness is immediately apparent, as estimates under this model exhibit high levels of reliability nationwide, despite known geographic and between-group disparities in the $y_{ik}$ (as shown in Table~\ref{tab:hd_desc_table}).
%As we can see, the standard model produces high levels of reliability nationwide due to the excessive informativeness of the model. For example, in the case of Black females (Figure~\ref{fig:rel_bf}), many areas achieve greater than 95\% level of reliability despite Table~\ref{tab:hd_desc_table} indicating the limited amount of data in most counties.
In contrast, the restricted MCAR model produces estimates whose levels of reliability are commensurate with the amount of data (as previously highlighted in Figure~\ref{fig:rp}).
%under the restricted model (Figure~\ref{fig:rel_res_bf}), the reliability of the estimates is substantially lower, aligning with the fact that most areas, except for coastal regions and the Southeast, have insufficient data. This pattern is not limited to Black females, who have the least available information. Even for White males (Figure~\ref{fig:rel_wm}), where almost all areas are classified as reliable, about 67\% of counties had fewer than 10 deaths, and 7\% had no observed deaths. Although this group has the most overall information, the assumption of high reliability across all counties is unrealistic, particularly in states with lower populations, such as North and South Dakota. However, under the restricted model, the estimated rates better reflect the geographical distribution of available information (Figure~\ref{fig:rel_res_wm}).

\section{Conclusion and Discussion} \label{sec:conclusion}
As discussed in Section~\ref{sec:intro}, there were two primary goals of this work.  First and foremost, a goal of this work was to provide a means by which users of the MCAR model could quantify the model's informativeness.  While \citet{bym:info} discussed how creating relative precision plots like those in Figure~\ref{fig:rp} could serve as an empirical, \emph{ad hoc} means by which to evaluate the informativeness of \emph{any} disease mapping model, the work presented in Section~\ref{sec:methods} provides a framework for researchers to \emph{calculate} the informativeness of the MCAR model.  We believe this is valuable in-and-of-itself because it provides transparency and because it can require researchers to defend or rethink their model specification when $\widehat{a}_{0k}$ is high.
%---when $\widehat{a}_k$ is high, as observed in Figure~\ref{fig:info_ci}---

Assuming that most disease mapping researchers would find the $\widehat{a}_{0k}$ levels in Figure~\ref{fig:info_ci} indefensibly (if not \emph{unacceptably}) high, the natural next question would be ``\emph{What can be done about it?}''  While one approach could be to modify the hyperparameters used in the priors for $\tau_k^2$ and/or $\bSig$ to encourage more uncertainty, heterogeneity across maps and different outcomes would likely require tailored prior specifications to yield consistent levels of $\widehat{a}_{0k}$.  As such, this is where the second primary goal of this work---developing a strategy for restricting the model's informativeness---comes into play.  Specifically, allowing users to directly restrict the model's informativeness to a defensible range of values allows them to focus their time and attention to more important aspects of their research.  That said, users \emph{would} need to define what a ``defensible'' range for $\widehat{a}_{0k}$ is for them; this is where the reliability criteria of \citet{quick:reliable} may offer a holistic way to view the problem.

%Our work extends prior studies on evaluating model informativeness from univariate to multivariate {\color{blue}spatial} settings. Assessing informativeness in the multivariate context is critical as it highlights the balance between borrowing strength across groups and preventing the model from overly dominating the data. Importantly, in relation to reliability, excessive informativeness may result in overly precise estimates, potentially leading to falsely reliable conclusions even in the absence of sufficient data. To prevent oversmoothing and properly assess certainty, we recommend that researchers conducting spatial analyses to evaluate the informativeness of their models and relative precision of their estimating and---if possible---limit the model's informativeness to enable fair comparisons across groups and time periods.

%In the MCAR model, to achieve the restriction using each group’s conditional distribution, we formulated the MCAR model to decompose into a series of marginal and conditional distributions, which we termed as ``regression approach.'' Previous work by \citet{jin:Generalized2005} used a similar approach in a bivariate spatial {\color{blue}setting} suggested that group ordering might affect outcomes. However, in our model, by using the conjugate inverse Wishart prior and decomposing it to facilitate our parameterization, we observed negligible differences in output across orderings.

As discussed in Section~\ref{sec:analysis}, an inverse Wishart prior was used in our analysis to ensure that the ordering of our groups did not impact the results.  While this %``order-agnostic''
feature is attractive, the flexibility of the ``regression approach'' could allow us to avoid some of the limitations associated with the inverse Wishart prior noted by other researchers, such as the fact that it %. For instance, the inverse Wishart prior
is controlled by a single degree of freedom for all components of the covariance matrix, often failing to capture its true structure \citep{barnard:Modeling2000}. In our future work, we intend to explore the viability of implementing a restricted MCAR using alternative prior specifications for covariance matrices---e.g., the scaled inverse Wishart of \citet{omalley:DomainLevel2008}, which separates the marginal variance parameters from the correlation parameters in $\bSig$.  If these fail to strike a balance between model flexibility and computational efficiency and convenience, we could also consider approaches that directly specify priors on the sequential conditional variance parameters, $\sig_{k \given <k}^2$, and regression coefficients, $\bbeta_k$, that aim to maintain the ``order-free'' nature of the inverse Wishart structure.
%By decomposing the covariance matrix into $\sig^2_{k \given < k}$ and $\bbeta_{k}$, we can apply separate inverse gamma and normal priors within a Gibbs sampler, without requiring these priors to match the structure of the inverse Wishart setup. This may provide a better representation of covariance structure, especially in cases of sparse data.

Finally, we note that in addition to the group-level dependencies modeled here, it is also possible to use MCAR models to incorporate yearly autocorrelation. This could help capture better trajectories of mortality rates by allowing the model to borrow information from previous years. However, to maintain interpretability and parsimony, one could consider alternative covariance structures, such as an AR(1) structure, in the restricted MCAR model. The regression approach naturally accounts for both marginal and conditional structures of multiple years given the natural ordering of the data. We expect our approach will be adaptable to different covariance structures, helping to model dependencies while seeking a balance between the model and data to ensure confidence in rate estimate reliabilities.

% \section*{Disclosure Statement}

% The authors declare no competing interests.

%\newpage
\bibliographystyle{LaTeX_files/Jasa}
\bibliography{LaTeX_files/PhD}

\end{document}